\documentclass[journal,twoside]{IEEEtran}
\usepackage{graphicx}
\usepackage{multirow}
\usepackage{textcomp}
\usepackage{subcaption}
\usepackage{amsmath,amssymb,amsfonts}
\usepackage{amsmath}
\usepackage{breqn}
\usepackage{mathtools}
\usepackage{lipsum}
\usepackage{pdflscape}
\usepackage{tabularx,booktabs, longtable}
\usepackage{diagbox}
\usepackage{array}
\usepackage{makecell}
\usepackage{bm}
\usepackage{adjustbox}
\usepackage{float}
\usepackage{scalerel,stackengine}
\stackMath
\newcommand\reallywidehat[1]{%
\savestack{\tmpbox}{\stretchto{%
  \scaleto{%
    \scalerel*[\widthof{\ensuremath{#1}}]{\kern-.6pt\bigwedge\kern-.6pt}%
    {\rule[-\textheight/2]{1ex}{\textheight}}%WIDTH-LIMITED BIG WEDGE
  }{\textheight}% 
}{0.5ex}}%
\stackon[1pt]{#1}{\tmpbox}%
}

\begin{document}

\title{Detection of Epileptic Seizures on EEG Signals Using ANFIS Classifier, Autoencoders and Fuzzy Entropies}

\author{Afshin~Shoeibi,
        Navid~Ghassemi,
        Marjane~Khodatars,
        Parisa~Moridian,
        Roohallah~Alizadehsani,
        Assef~Zare,
        Abbas~Khosravi,
        Abdulhamit~Subasi,
        U.~Rajendra~Acharya
        and~J.~Manuel~Gorriz% <-this % stops a space
 \thanks{ A. Shoeibi and N. Ghassemi are with the Faculty of Electrical Engineering, Biomedical Data Acquisition Lab (BDAL), K. N. Toosi University of Technology, Tehran, Iran, Iran. (Corresponding Author: Afshin Shoeibi, e-mail: afshin.shoeibi@gmail.com).}
 \thanks{ M. Khodatars is with the Department of Medical Engineering, Mashhad Branch, Islamic Azad University, Mashhad, Iran.}
 \thanks{ P. Moridian is with the Faculty of Engineering, Science and Research Branch, Islamic Azad University, Tehran, Iran.}%
 \thanks{ R. Alizadehsani and A. Khosravi are with the Institute for Intelligent Systems Research and Innovation (IISRI), Deakin University,Victoria 3217, Australia.}%
 \thanks{ A. Zare is with the Faculty of Electrical Engineering, Gonabad Branch, Islamic Azad University, Gonabad, Iran.}
 \thanks{ A. Subasi is with the Institute of Biomedicine, Faculty of Medicine, University of Turku, 20520, Turku, Finland. Also with the Department of Computer Science, College of Engineering, Effat University, Jeddah, 21478, Saudi Arabia.}
 \thanks{ U. R. Acharya is with the Department of Electronics and Computer Engineering, Ngee Ann Polytechnic, Singapore 599489, the Department of Biomedical Engineering, School of Science and Technology, SUSS University, Singapore 599491 and the Department of Biomedical Informatics and Medical Engineering, Asia University, Taichung, Taiwan.}
 \thanks{ Juan M. Gorriz is with the Department of Signal Theory, Networking and Communications, Universidad de Granada, Spain. Also with the Department of Psychiatry. University of Cambridge, UK.}%
% <-this % stops a space
% \thanks{Manuscript received April 19, 2005; revised August 26, 2015.}
}

\maketitle

\begin{abstract}
Epileptic seizures are one of the most crucial neurological disorders, and their early diagnosis will help the clinicians to provide accurate treatment for the patients. The electroencephalogram (EEG) signals are widely used for epileptic seizures detection, which provides specialists with substantial information about the functioning of the brain. In this paper, a novel diagnostic procedure using fuzzy theory and deep learning techniques is introduced. The proposed method is evaluated on the Bonn University dataset with six classification combinations and also on the Freiburg dataset. The tunable-Q wavelet transform (TQWT) is employed to decompose the EEG signals into different sub-bands. In the feature extraction step, 13 different fuzzy entropies are calculated from different sub-bands of TQWT, and their computational complexities are calculated to help researchers choose the best set for various tasks. In the following, an autoencoder (AE) with six layers is employed for dimensionality reduction. Finally, the standard adaptive neuro-fuzzy inference system (ANFIS), and also its variants with grasshopper optimization algorithm (ANFIS-GOA), particle swarm optimization (ANFIS-PSO), and breeding swarm optimization (ANFIS-BS) methods are used for classification. Using our proposed method, ANFIS-BS method has obtained an accuracy of 99.74% in classifying into two classes and an accuracy of 99.46% in ternary classification on the Bonn dataset and 99.28% on the Freiburg dataset, reaching state-of-the-art performances on both of them.
\end{abstract}


\IEEEpeerreviewmaketitle

\section{Introduction}

\IEEEPARstart{E}{pilepsy} is a group of neurological disorders associated with seizures \cite{one,oner}. Epileptic seizures result from the abnormal activity in the cortex of the brain, presented in two types of focal and general seizures \cite{added1,addedr1}. Seizures can cause various symptoms such as sudden loss of consciousness, muscle contraction, and emotional or behavioral changes in the patient \cite{ttwo_ei,ttwo_eier}. Based on statistics published by the world health organization (WHO), more than 50 million people on the planet suffer the burden of epilepsy, shaping around one percent of the world population \cite{riper1}.

Neuroimaging modalities are among the essential techniques for epileptic seizures detection, including functional and structural methods \cite{ntwo2,ntwo3}.

The EEG signals carry vital physiological and pathological information in their recordings \cite{two}. Various neurological disorders such as epileptic seizures \cite{three} and autism spectrum disorder (ASD) \cite{four} can be diagnosed using EEG signals, owing to its high temporal resolution, portability, and inexpensiveness. Epileptic seizures are detected currently by clinicians by visual inspection, which is subjective and time-consuming. The presence of myogenic and ocular artifacts makes the detection process a challenging task \cite{five}. The computer aided diagnosis system (CADS) can be employed for the diagnosis of epileptic seizures. The CADS helps clinicians to identify epileptic seizures automatically \cite{six}. These systems are faster, more accurate, and also can handle a huge volume of data. The CADS has four steps for automated diagnosis of epileptic seizures. These steps contain pre-processing, features extraction, feature selection/dimensionality reduction, and classification \cite{one,added1,addedr1,ttwo_ei,ttwo_eier,seven}.

\begin{figure*}[t]
    \centering
    \includegraphics[width=6.5in]{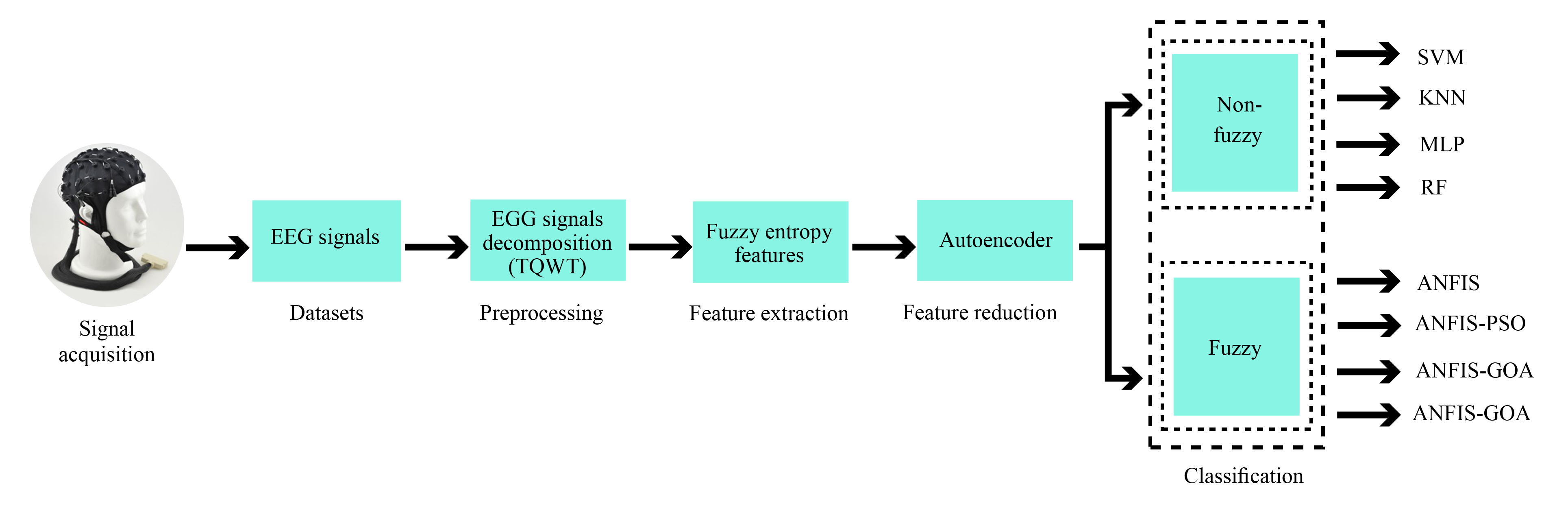}
    \caption{Proposed method for automated diagnosis of epileptic seizures.}
    \label{fig:one}
\end{figure*}

So far, many pieces of research have been conducted on creating CADS suitable for epileptic seizures detection on EEG signals, mostly focusing on increasing these systems' accuracy and performance. In this paper, a novel method is presented for epileptic seizure detection on EEG signals based on fuzzy logic theories and deep learning; Figure \ref{fig:one} shows the steps of the proposed system. As shown in Figure \ref{fig:one}, the Bonn and the Freiburg datasets were used for the implementation and evaluation of the proposed method. The Bonn dataset contains various classification problems.

The second part of the proposed CADS is dedicated to preprocessing EEG signals of both datasets using TQWT. In this stage, the TQWT \cite{two_eight} is used to decompose the EEG signals of the Bonn \cite{two_nine} and Freiburg \cite{nb1} datasets into different sub-bands. TQWT, first introduced in \cite{two_eight}, is an improved version of DWT, suitable for preprocessing of chaotic biomedical signals such as EEG. Redundancy or oversampling rate ($r$), number of sub-bands ($J$), and Q factor ($Q$) are the most critical parameters of this wavelet transform. In our paper, parameters are chosen as $J=8$, $r=1$, and $Q=3$, similar to \cite{eight}.

Considering that EEG signals have chaotic behavior, non-linear feature extraction techniques can help CADS reach higher performances. Feature extraction is of utmost importance in steps of CADS for epileptic seizure detection. Previous studies have shown that non-linear features perform better in this task, given the chaotic characteristics of EEG signals \cite{ent}. Among them, fractal dimension (FD) techniques \cite{fractal}, correlation \cite{cor}, largest Lyapunov exponent (LLE) \cite{llp}, and also various types of entropies \cite{ent}.

In this paper, a combination of fuzzy entropies is employed for feature extraction. Employing fuzzy entropies for epileptic seizure detection has been examined very limited in previous studies, all of which are reviewed in the following.

Xing et al. \cite{new1} proposed a method for epileptic seizure detection based on fuzzy entropy (FuEn). They used FuEn for feature extraction and support vector machine (SVM) for classification and achieved promising results. In \cite{new2}, first, EEG signals were segmented into various time-windows. In the following, power spectral density (PSD) and FuEn were computed for signals as features; lastly, SVM was applied for classification, and an accuracy of 96.20\% is reached at best. Tripathi et al. \cite{new3} employed empirical mode decomposition (EMD) in preprocessing step for the decomposition of signals into different sub-bands; then, they used FuEn and SVM for feature extraction and classification, respectively, and reported a 97\% accuracy. In \cite{two_five}, the authors introduced a novel entropy named fuzzy distribution entropy (FuDistEn). Then, an epileptic seizures detection method is presented that consists of wavelet packet decomposition (WPD), FuDistEn, and Kruskal-Wallis nonparametric one-way analysis of variance. In their system, FuDistEn values are computed from all nodes in every level and fed to k nearest neighbors (KNN) classifier as features.

An epileptic seizure detection method based on fuzzy permutation entropy (FuPeEn) was introduced in \cite{new4}. In this work, an artificial neural network (ANN) is employed for classification, and an accuracy of 98.72\% is achieved. Similarly, in \cite{three_nine}, FuPeEn has been picked for the same task. Here, after preprocessing the input signals, researchers have extracted FuPeEn from each frame and fed them to various classification methods for comparison. Averagely, they reach an accuracy of 98.72\%. In another research, Kumar et al. \cite{new5} applied fuzzy approximate entropy (FuApEn) for epileptic seizure detection. They first preprocessed signals using discrete wavelet transform (DWT), and after feature extraction, they used an SVM for classification, reporting an accuracy of 97.83\%. Bhattacharyya et al. \cite{new7} employed TQWT and multivariate Fuzzy Entropy for preprocessing and feature extraction, respectively. Fractional fourier transform-wavelet packed transform (FFTWPT) was used by Li et al. \cite{new8} in preprocessing step of their method. Then, FuEn was used for feature extraction. By using principal component analysis (PCA) for dimensionality reduction and SVM for classification, they reached an accuracy of 98.58\%.

As noticeable from reviewed studies, no previous work has been conducted on combining various Fuzzy entropies as feature extractors from EEG signals for epileptic seizures detection. In this paper, numerous fuzzy entropy based features are extracted from different sub-bands of TQWT for epileptic seizure detection, which makes this paper also a suitable review of type-1 fuzzy entropy features for this task. The different fuzzy entropies are FuEn \cite{three_zero}, averaged (AFuEn) \cite{three_one}, multiscale (MFuEn) \cite{three_two}, refined composite multiscale (RCMFuEn) \cite{three_three}, fractional (FFuEn) \cite{three_four}, FuApEn \cite{three_five}, minimum variance modified (MVMFuEn) \cite{three_six}, inherent (IFuEn) \cite{three_seven}, FuDistEn \cite{two_five}, cross fuzzy (CFuEn) \cite{three_eight}, FuPeEn \cite{three_nine}, hierarchical (HFuEn) \cite{four_zero}, and fuzzy measure (FuMeEn) \cite{four_one}.

As demonstrated in Figure \ref{fig:one}, the fourth step is dedicated to dimensionality reduction. In this paper, an AE is applied to reduce the feature matrix's dimensionality; the applied AE has a hand-tailored structure for this specific task, which is the second novelty of this paper. Considering the high dimensionality of feature vectors, without dimensionality reduction, the classifier cannot train properly; AE is a well-known dimensionality reduction that learns a representation in a smaller space. Compared to other dimensionality reduction methods such as PCA \cite{asvm}, AEs are capable of modeling complex non-linear functions, while others are mostly a linear transformation.

The last step of the proposed method is classification; here, KNN, multilayer perceptron (MLP), SVM, random forest (RF), and various ANFIS classification methods are employed for comparison. First, the classification is performed using standard ANFIS \cite{four_four}; then, to improve the proposed CADS performance, improved ANFIS-GOA, ANFIS-PSO, and ANFIS-BS models are used. Standard ANFIS uses backpropagation to train the model; however, in the improved versions, GOA \cite{goanew}, PSO \cite{five_two}, and BS \cite{five_three} methods are used to enhance the training process. Using optimizers for training the ANFIS has been studied before by many researchers \cite{anfga,anfpso}; however, they have mostly used GA or PSO optimizers. In this work, we have used the BS algorithm to train the ANFIS, which is a combination of GA and PSO, and to the best of our knowledge, this is the first time any research has used this algorithm to train ANFIS. Results show the superiority of the proposed method compared to previous studies.

The organization of the rest of the paper is depicted as follows. The material and methods are described in Section 2, Section 3 is dedicated to the evaluation processes; and finally, Section 4, 5, and 6 present the experiment results, limitations of the study, and discussion, respectively.

\section{Material and Methods}
\subsection{Dataset}

\subsubsection{Bonn Dataset}

This dataset was recorded at the University of Bonn by a group of researchers, and it had been extensively used for research in the area of epileptic seizure analysis and detection \cite{two_nine}. This dataset contains 500 signal frames with a length of 23.6 seconds. In this dataset, the sampling frequency is 173.61~Hz. They consisted of five classes' viz. A, B, C, D, and E with 100 segments recordings in each class \cite{two_nine}. Five healthy controls in the relaxed and awake state with 10-20 standard electrode placement schemes contributed to Class A and B EEG surface data. Intracranial electrodes were used on 5 patients who suffered from epilepsy and collected data of C, D, and E classes \cite{two_nine}.

The hemisphere of the epileptogenic zone and the opposite hemisphere were used respectively for the recording of the C and D classes’ signals during inter-ictal (seizure-free) period. Ictal (seizure) period was taken into account for the recording of class E \cite{two_nine}. More details about the Bonn dataset are reported in Tables \ref{tab:table_one} and \ref{tab:table_two}. 
% Also, a schematic representation of each group of this dataset is shown in Figure \ref{fig:three}.

\subsubsection{Freiburg Dataset}
The Freiburg dataset \cite{nb1} includes intracranial electroencephalography (IEEG) signals from 21 patients suffering from focal epileptic seizures recorded at Freiburg Hospital in Germany. All IEEG signals were recorded using a Neurofile NT digital video EEG system with 256~Hz sampling frequency. During recording, depth (d), strip (s), and grid (g) electrodes were applied to reduce noise and increase SNR. In this database, all subjects range in age from 10 to 50 years containing 13 women and 8 men. The database comprises a variety of ictal, pre-ictal, and inter-ictal samples from patients with epileptic seizures. At the time of recording IEEG signals, at least 2 and at most 5 epileptic seizures were observed in each subject. Three different seizure types including generalized tonic-clonic (GTC), complex partial (CP), and simple partial (SP) have been reported among patients and they have experienced at least two different types. Epilepsy was detected in 11 patients in the neocortical brain location, in the hippocampus location in eight patients, and in two patients in both regions. More details on Freiburg dataset are provided in Table \ref{tab:fry}.
\begin{table}[t]
    \centering
    \caption{\textbf{Thorough explanation of five subsets of dataset}}
    \resizebox{ \linewidth }{!}{
    \begin{tabular}{cccccc}
        \Xhline{2\arrayrulewidth}
        \hline
        \hline
        \Xhline{2\arrayrulewidth}
        \multirow{2}{*}{\shortstack[1]{\thead{Sets}}} & \multicolumn{5}{c}{\thead{Subjects}} \\
        \cline{2-6} & \thead{Patient \\ Stage} & \thead{Electrode \\ type} & \thead{Num. of \\ Cases} & \thead{Num. of \\ Data} & \thead{Length. of \\ Segments} \\
        \Xhline{2\arrayrulewidth}
        Set A & Eye Open            & Surface      & 5 & 100 & 4097 \\
        Set B & Eye Close           & Surface      & 5 & 100 & 4097 \\
        Set C & Seizure Free        & Intracranial & 5 & 100 & 4097 \\
        Set D & Seizure Free        & Intracranial & 5 & 100 & 4097 \\
        Set E & Seizure Activity    & Intracranial & 5 & 100 & 4097 \\
        \Xhline{2\arrayrulewidth}
        \hline
        \hline
        \Xhline{2\arrayrulewidth}
    \end{tabular}}
    \label{tab:table_one}
\end{table}

\begin{table*}[ht]
    \centering
    \caption{\textbf{More details on the Freiburg dataset}}
    \resizebox{0.85\linewidth }{!}{
    \begin{tabular}{cccccccc}
        \Xhline{2\arrayrulewidth}
        \hline
        \hline
        \Xhline{2\arrayrulewidth}
\thead{Patient} & \thead{Age} & \thead{Gender} & \thead{Seizure Origin} & \thead{Seizure Type} & \thead{Electrodes} & \thead{Number of Seizures} & \thead{Focus Location}\\
\Xhline{2\arrayrulewidth}
1 & 15  & Female & Temporal & SP, CP & g, s & 4 & Ne \\
2 & 38  & Male   & Frontal & SP, CP, GTC  & d & 3 & Hi \\
3 & 14  & Male   & Temporal & SP, CP & g, s & 5 & Ne \\
4 & 26  & Female & Temporal & SP, CP, GTC  & d, g, s & 5 & Hi \\
5 & 16  & Female & Frontal & SP, CP, GTC  & g, s & 5 & Ne \\
6 & 31  & Female & Temporal & CP, GTC & d, g, s & 3 & Hi \\
7 & 42  & Female & Temporal & SP, CP, GTC  & d & 3 & Hi \\
8 & 32  & Female & Temporal & SP, CP & g, s & 2 & Ne \\
9 & 44  & Male & Frontal & CP, GTC & g, s & 5 & Ne \\
10 & 47 & Male & Frontal & SP, CP, GTC  & d & 5 & Hi \\
11 & 10  & Female & Frontal & SP, CP, GTC  & g, s & 4 & Ne \\
12 & 42  & Female & Frontal & SP, CP, GTC  & d, g, s & 4 & Hi \\
13 & 22  & Female & Temporal & SP, CP, GTC  & d, s & 2 & Hi \\
14 & 41  & Female & Temporal & CP, GTC & d, s & 4 & Ne, Hi \\
15 & 31  & Male   & Frontal & SP, CP, GTC  & d, s & 4 & Ne, Hi \\
16 & 50  & Female & Temporal & SP, CP, GTC & d, s & 5 & Hi \\
17 & 28  & Male & Temporal & SP, CP, GTC & s & 5 & Ne \\
18 & 25  & Female & Temporal & SP, CP & s & 5 & Ne \\
19 & 28  & Female & Frontal & SP, CP, GTC & s & 4 & Ne \\
20 & 33  & Male   & Temporal & SP, CP, GTC & d, s & 5 & Ne \\
21 & 13  & Male & Temporal & SP, CP & s & 5 & Ne\\
\Xhline{2\arrayrulewidth}
        \hline
        \hline
        \Xhline{2\arrayrulewidth}
    \end{tabular}}
    \label{tab:fry}
\end{table*}

\subsection{Preprocessing using TQWT}

\begin{table}[t]
    \centering
    \caption{\textbf{More details about six problem classification}}
    \resizebox{ \linewidth }{!}{
    \begin{tabular}{ccc}
        \Xhline{2\arrayrulewidth}
        \hline
        \hline
        \Xhline{2\arrayrulewidth}
        \thead{Case} & \thead{Classification \\ Problems} & \thead{Description}\\
        \Xhline{2\arrayrulewidth}
        Case 1 & A-E & Healthy - Ictal \\
        Case 2 & B-E & Healthy - Ictal \\
    
        Case 3 & C-E & Interictal - Ictal \\
        
        Case 4 & D-E & Interictal - Ictal \\
        
        Case 5 & ABCD-E & Normal - Seizure \\
        
        Case 6 & AB-CD-E &  Healthy - Interictal - Seizure \\
        \Xhline{2\arrayrulewidth}
        \hline
        \hline
        \Xhline{2\arrayrulewidth}
    \end{tabular}}
    \label{tab:table_two}
\end{table}

Wavelet transforms have a wide range of applications in the scope of brain signal processing \cite{aw3,aw4,aw5}. TQWT is an improved and particular type of DWT introduced by Ivan Selesnick \cite{two_eight} and is also employed to precisely analyze EEG signals. In the TQWT, the input parameters, $r$, $J$, and $Q$ can be tuned \cite{two_eight}. Two channel filter bank operations high pass filter $H_1(\omega)$ with scaling factor $\gamma$ and $\xi$ low pass filter $H_0(\omega)$ with scaling factor $\xi$ can be used in TQWT. The TQWT expressions are as given below \cite{two_eight}: 
\begin{equation} \label{eq:one}
H_0(\omega)= \left\{ 
\begin{array}{lr}
1 & |\omega|<(1-\gamma)\pi\\
\lambda(\frac{\omega+(\gamma - 1)\pi}{\xi+\gamma-1}) & (1-\gamma)\pi \le |\omega| < \xi \pi\\
0 & \xi\pi\le|\omega|<\pi
\end{array}
 \right. 
\end{equation}

\begin{equation}
\label{eq:two}
H_1(\omega)= \left\{
\begin{array}{lr}
0&|\omega|<(1-\gamma)\pi\\
\lambda(\frac{\xi\pi-\omega}{\xi+\gamma-1}) & (1-\gamma)\pi \le |\omega| < \xi \pi \\
1 & \xi\pi\le|\omega|<\pi
\end{array}
\right. 
\end{equation}

where $\lambda(\omega)=\cos^2 \frac{\omega}{2} \sqrt{2-\cos (\omega)},|\omega|\le\pi$ is the Daubechies filter frequency response \cite{two_eight}.

High-pass and low-pass scaling factors are selected to satisfy the conditions \cite{two_eight}:
\begin{equation}
\label{eq:three}
    0<\xi<1;0<\gamma\le1;\xi+\gamma>1
\end{equation}

The maximum number of sub-bands $J_{\text{max}}$, redundancy $r$, and quality factor $Q$ and  parameters are defined in terms of $\xi$ and $\gamma$ as \cite{two_eight}: 
\begin{equation}
\label{eq:four}
    r=\frac{\gamma}{1-\xi};Q=\frac{2-\gamma}{\gamma};J_{\text{max}}=\frac{\log (\frac{\gamma^N}{8})}{\log (\frac{1}{\xi})}
\end{equation}

\begin{figure}[t]
    \centering
    \includegraphics[width=3.4in]{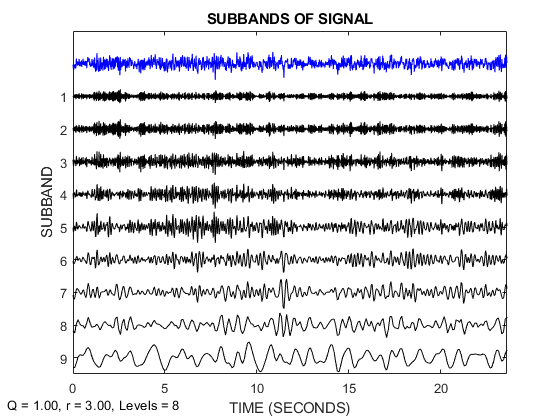}
    
    \caption{EEG signal of Bonn dataset decomposition using TQWT.}
    \label{fig:two}
\end{figure}

\begin{figure}[t]
    \centering
    \includegraphics[width=3.4in]{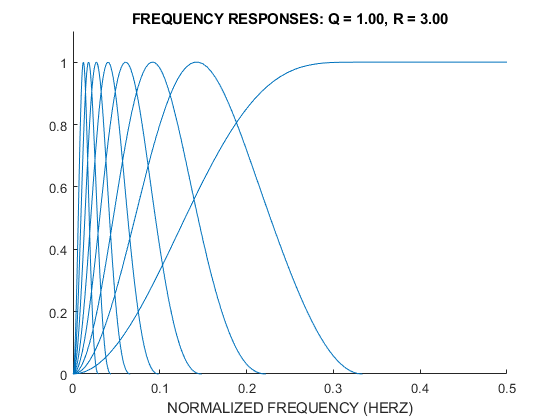}
    
    \caption{Illustration of frequency response for r=3, Q=1, and J=8.}
    \label{fig:freq_rep}
\end{figure}

Redundancy r must be greater than 1, and the Q-factor should be chosen such that $Q \le$ 1. Figure \ref{fig:two} illustrates the result of applying TQWT on a frame of the Bonn dataset's signals for $r=3$, $Q=1$, $J=8$, respectively. Also, Figure \ref{fig:freq_rep} illustrates the frequency response for $r=3$, $Q=1$, and $J=8$.

\subsection{Feature Extraction}
EEG signals have chaotic behavior. Given that entropies are an important class of feature extraction methods, in this work, different fuzzy entropies have been used to extract the feature from EEG signals, and they are mentioned below.

\subsubsection{Standard Fuzzy Entropy}
For a time series $x(i),i=\text{1,2,...,}N$ FuEn \cite{three_zero} establishes vector sequences $\bm{X}^{m}_{i},i=1,...,N-m+1$ as given blow:

\begin{equation}
\label{eq:five}
    \bm{X}_i^m=\{x(i),x(i+1),...,x(i+m-1)\}-x_{0}(i)
\end{equation}

Where the length of the sequences is denoted by $m$, $x_{\text{0}}(i)$ is a baseline. 

$\text{D}_{ij}^m(n,r)$ is the similarity degree using fuzzy membership function $\mu(d_{ij}^m,n,r))$ for the vector $X_i^m$ and $X_j^m$ replacing the Heaviside function \cite{three_zero}.

\begin{equation}
\label{eq:six}
    D_{ij}^m(n,r)=\mu(d_{ij}^m,n,r)
\end{equation}

\begin{equation}
\label{eq:seven}
   \mu(d_{ij}^m,n,r)=e^{\frac{-(d_{ij}^m)^n}{r}}
\end{equation}

where r and n are predefined gradient and width of the exponential function, $d_{ij}^m$ is the maximum absolute difference between $\bm{X}_i^m$ and $\bm{X}_j^m$. $\phi^m $  function is defined as below \cite{three_zero}:
\begin{equation} 
\label{eq:eight}
    \phi^m(n,r)=\frac{1}{N-m} \sum_{i=1}^{N-m}\left(\frac{1}{N-m-1}\sum_{j=1,j \ne i}^{N-m} D_{ij}^m\right)
\end{equation}

The sequences $\bm{X}_{i}^{m+1}$ is generated by setting $m$ ← $m$+1 and $\phi^{m}(n,r)$ is constructed afterwards. Time series for input $x(i)$ for FuEn is generated $\phi^{m}(n,r)$ deviated from $\phi^{m+1}(n,r)$ as given below \cite{three_zero}:
\begin{equation} \label{eq:nine}
    \text{FuEn}(m,n,r,N) = \frac{\phi^{m}(n,r)}{\phi^{m+1}(n,r)}
\end{equation}
\subsubsection{Averaged Fuzzy Entropy}

AFuEn is a novel and improved model of FuEn. Various approaches have been proposed to implement this entropy \cite{three_one}. In this method, an improved m\_pattern $\Gamma_{k}[\bm{X}_j^m]$ is compared to $\bm{X}_i^m$ . At this entropy, equation (\ref{eq:six}) is modified as follows \cite{three_one}:
\begin{equation} \label{eq:ten}
    ^kD_{ij}^m(n,r)=\mu\left(d \left[ \bm{X}_i^m , \Gamma_k [\bm{X}_j^m]\right] , n,r \right)
\end{equation}
In the following, four different types of $\Gamma_k\left[\bm{X}_m(j)\right]$ operations with k=\{T,R,I,G\} are defined as follows \cite{three_one}:
\begin{itemize}
    \item A translation of $n$ samples, $k=T$ is corresponding to $\Gamma_T [ \bm{X}_j^m] = \bm{X}_{j+n}^m$ 
    \item A reflection at the position $n$, $k=R$ is corresponding to $\Gamma_R[\bm{X}_j^m]=\bm{X}_{-j+n}^m$
    \item An inversion at the position $n$, $k=I$ is corresponding to $\Gamma_I[\bm{X}_j^m]=-\bm{X}_{-j+n}^m$
    \item A glide reflection of $n$ samples, $k=G$ is corresponding to $\Gamma_G[\bm{X}_j^m]=-\bm{X}_{j+n}^m$
\end{itemize}

In this case, four fuzzy entropies including FuEn\textsubscript{T}, FuEn\textsubscript{R}, FuEn\textsubscript{I}, and finally FuEn\textsubscript{G} are obtained. The following fuzzy entropy is computed as follows \cite{three_one}:
\begin{equation} \label{eq:one_one}
    \text{FuEn}_a(m,n,r,N)=\frac{(\text{FuEn}_T + \text{FuEn}_R + \text{FuEn}_I + \text{FuEn}_G)}{4}
\end{equation}
Finally, equation (\ref{eq:one_two}) is acquired.
\begin{equation}
\label{eq:one_two}
    \text{FuEn}_k(m,n,r,N)=\ln{\left[\frac{\phi_k^m(n,r)}{\phi_k^{m+1}(n,r)}\right]}
\end{equation}

\subsubsection{Multi-Scale Fuzzy Entropy}
In MFuEn, slight variations in EEG signals are detected with high precision. At this entropy, we have \cite{three_two}:
\begin{equation} \label{eq:ur1}
    y_{\tau}(j)=\frac{1}{\tau} \sum_{i=(j-1)\tau+1}^{j\tau} u_i;i \le j\le \frac{N}{\tau}
\end{equation}

Where $y_{\tau}(j)$ represents the large series of the structure, $\tau$ is the scale factor, $u_i$ indicates the independent components of the time series, and $N$ is the total data points in the EEG signals. Finally, MFuEn for any large series of the structure $y_{\tau}(j)$ is calculated using the equation below \cite{three_two}:
\begin{equation}\label{eq:ur2}
    \text{MFuEn}(m,r) = \log(\frac{B_r^m}{A_r^{m+1}})
\end{equation}
$A$ and $B$ are counters that analyze $m$ and $(m+1)$ pattern matchings with the value $r$.
\subsubsection{Refined Composite Multiscale Fuzzy Entropy}

RCMFuEn based on $\mu$ and $\sigma$ are the two techniques introduced by \cite{three_three}. For embedding dimension m, scale factor $\tau$, $\phi_{\tau,k}^m | (k = 1,...,\tau)$  and $\phi_{\tau,k}^{m+1}| (k = 1,...,\tau)$  for each $Z_k^{(r)} | (k = 1,...,\tau)$  are calculated separately. The $\text{RCMFuEn}\sigma$ is computed as follows \cite{three_three}: 

\begin{equation} \label{eq:one_four}
    \text{RCMFuE}\sigma(x,\tau,m,n,r)=-\ln{\left( \frac{\overline{\phi}_r^{m+1}}{\overline{\phi}_r^m} \right) }
\end{equation}

$\text{RCMFuEn}\sigma$ and $\text{RCMFuEn}\mu$ have differences that both use different equations in their first step algorithm. The tolerance $r$, Fuzzy entropy power $n$, the embedding dimension $m$ were chosen as 0.15, 2, and 2 respectively \cite{three_three}.

\subsubsection{Fractional Fuzzy Entropy}

As shown in \cite{three_four}, a new fractional based Shannon entropy is introduced as follows:
\begin{equation}
    S_{\alpha} = \sum_i p_i  \left[ -\frac{p_i^{-\alpha}}{\Gamma(\alpha + 1)} \left[ \ln p_i+ \psi(1) - \psi(1- \alpha) \right] \right]
\end{equation}

Also, fractional-order information of order $\alpha$ can be deducted using : 
\begin{equation}
I_\alpha = -\frac{p_i^{-\alpha}}{\Gamma(\alpha + 1)} [\ln p_i+ \psi(1) - \psi(1- \alpha)]
\end{equation}

Using the base idea of the Shannon fractional entropy, in \cite{three_four} a Fuzzy based fractional entropy is introduced, which can be shown similar to equation (\ref{eq:x}).
\begin{equation}\label{eq:x}
\begin{split}
&\text{FFuEn}(m, r, \alpha, x^N)=\\
&- \left( \frac{\phi^{m+1}(r)}{\phi^{m}(r)} \right)^{-\alpha} \frac{ln \frac{\phi^{m+1}(r)}{\phi^{m}(r)} + \psi(1) - \psi(1-\alpha)}{\Gamma(1+\alpha)}
\end{split}
\end{equation}

\subsubsection{Fuzzy Approximate Entropy}

In FuApEn, the similarity index depends on the fuzzy membership function \cite{three_five}. Hard boundaries are loosened by choosing the points approach over the Heaviside function. Similarity degree $\text{D}_{ij}^m$ between two vectors $\bm{X}_i^m$ and $\bm{X}_j^m$ is calculated by a fuzzy membership function by using a predetermined tolerance value $r$ as below \cite{three_five}:

\begin{equation} \label{eq:one_six}
    D_{ij}^m = \mu (d_{ij}^m,r)
\end{equation}

The function $\text{C}_r^m$  is given by

\begin{equation} \label{eq:one_seven}
    C_r^m(i) = \frac{1}{N-m+1}\sum_{j=1,j\ne i}^{N-m+1}D_{ij}^m
\end{equation}

\begin{equation} \label{eq:one_eight}
    \phi^m(r) = \frac{1}{N-m+1} \sum_{j=1,j\ne i}^{N-m+1} \ln [C_r^m(i)]
\end{equation}

$\text{FuApEn}(m, r)$ can be defined as time series measure from the function $\phi^{(m+1)}(r)$ and the vector sequence $\{ \bm{X}_i^{m+1} \}$ where sequence length $N$, tolerance $r$, and dimension $m$ as given below \cite{three_five}:

\begin{equation} \label{eq:one_nine}
    \text{FuApEn}(m,r)= \lim_{m\rightarrow \infty} [\phi^m(r)-\phi^{m+1}(r)]
\end{equation}

FuApEn can be defined for finite datasets from the statistics as below \cite{three_five}:

\begin{equation} \label{eq:two_zero}
    \text{FuApEn}(m,r,N) = \phi^m(r)-\phi^{m+1}(r)
\end{equation}

\subsubsection{Minimum Variance Modified Fuzzy Entropy}
The modified FuEn is proposed (MoFuEn) by \cite{three_six} to tackle membership functions limitations. The relative energy function $\psi$ is estimated by using $N$ as segmentation length, time-series $x={x_1,x_2,...,x_N}$ \cite{three_six}.

\begin{equation} \label{eq:two_one}
    \psi_i = \frac{|x_i|^2}{\sum_{i=1}^N|x_i|^2}
\end{equation}
The MoFuEn is calculated using relative energy as membership function as follows \cite{three_six}:
\begin{equation} \label{eq:two_two}
    H_A = -C\sum_{i=1}^N \left[ \psi_i\log (\psi_i) + (1 - \psi_i)\log(1 - \psi_i) \right]
\end{equation}
Let $H=H_{A_1},H_{A_2},...,H_{A_K}$ where $k$ is the total of modified fuzzy entropy values, then \cite{three_six}

\begin{equation} \label{eq:two_three}
    H_{\text{new}} = \frac{H}{\text{min}(H)}
\end{equation}
And MVMFuEn is defined as \cite{three_six}:
\begin{equation} \label{eq:two_four}
    \text{MVMFuEn}=\frac{H_{\text{new}}}{\sigma^2_{H_{\text{new}}}}
\end{equation}
Where $\sigma_{H_{\text{new}}}^2$ is the variable of $H_{\text{new}}$. More information about MVMFuEn is presented in \cite{three_six}.

\subsubsection{Inherent Fuzzy Entropy}
IFuEn is presented for the first time in \cite{three_seven} and includes three primary steps. In the first step, several IMFs are generated by decomposing the original signal $x(t)$ and regeneration of signal $\hat{x}(t)$ using EMD techniques. In the second step, FuEn is applied to evaluate the complexity. Finally, the MFuEn version similar to relations (\ref{eq:ur1}) and (\ref{eq:ur2}) is used for the third stage. The details of this entropy are presented in \cite{three_seven}.

\subsubsection{Fuzzy Distribution Entropy}
This entropy has for steps reconstruction of state-space, distance matrix construction, similarity degree calculation, probability density estimation, and FuDistEn calculation. In reconstruction of state-space stage, baseline removal is done like FuEn as a first step and vectors of $N-m+1$ m-dimensional are created \cite{two_five}. In step two, Chebyshev distance is utilized in FuDistEn. In order to estimate similarity degree $D_{ij}^m$ exponential function and fuzzy membership functions are used \cite{two_five}. In probability density estimation step, the diagonal elements of $D_{ij}^m$ are not included to shun self-matching and vector $R^m$ is obtained by transforming the matrix. Finally, FuDistEn calculation is as eq. (\ref{eq:url3}) as follow \cite{two_five}: 
\begin{equation}\label{eq:url3}
    \text{FuDistEn}(M,m,n,r) = -\sum^M_{t=1}P_t\frac{\log_2P_t}{\log_2M}
\end{equation}
Which $P_t$ be the probabilities of ePDF, tolerance $r$, the order of $n$, and embedding dimension $m$.

\subsubsection{Cross Fuzzy Entropy}

Two distinct signals’ similarity or synchrony can be analyzed using CFuEn \cite{three_seven}. CFuEn utilized exponential function instead of Heaviside function as used in cross-sample entropy for examining the similarity of vectors. The CFuEn parameters $m,n,r$ can be denoted as the negative logarithm of $\frac{\phi^{m+1}(n,r)}{\phi^{m}(n,r)}$ conditional probability \cite{three_eight}. 

\begin{equation} \label{eq:three_two}
    \text{CFuEn}(m,n,r,N)=-\lim_{N \rightarrow \infty}\left(\ln\left(\frac{\phi^{m+1}(n,r)}{\phi^{m}(n,r)}\right)\right)
\end{equation}
\subsubsection{Fuzzy Permutation Entropy}

FuPeEn is computed as follows \cite{three_nine}:
\begin{enumerate}
    \item Let us assume a time series $[x(i): 1 \le i \le L]$, where $L$ is the length of the series $\bm{X}$. These time series are utilized to create a matrix \cite{three_nine}.
    \begin{equation*} \label{eq:three_three}
        \begin{bmatrix}
        x(1) & x(1+\tau) & ... & x(1+(pm-1)\tau) \\
        x(2) & x(2+\tau) & ... & x(2+(pm-1)\tau)\\
        x(3) & x(3+\tau) & ... & x(2+(pm-1)\tau)\\
        ... & ...& ...&... \\
        x(j) & x(j+\tau) & ... & x(j+(pm-1)\tau)\\
        ... & ...& ...&... \\
        x(k) & x(k+\tau) & ... & x(k+(pm-1)\tau)
        \end{bmatrix}
    \end{equation*}
    where $pm$ and $\tau$ are the permuted dimension and the embedded time dimension, respectively. $k=L-(pm-1)\tau$ each row of the matrix is considered as a reconstruction component. Hence, there are $k$ reconstruction components in the matrix described above \cite{three_nine}.
    \item We can derive a new time series constructed from the earlier time series with the values between 1 and $pm!$ Arranging the elements in ascending order based upon their values \cite{three_nine}.
    \begin{equation} \label{eq:three_four}
        \left\{ U(i):1 \le i \le 1 - (pm - 1)\tau \right\}
    \end{equation}
    \item m-dimensional vector is constructed by considering the length of $U$ as $N$, by arranging the elements in order to reconstruct $U$ \cite{three_nine}.
    \begin{equation} \label{eq:three_five}
        Y_i^m=\{u(i),u(i+1),...,u(i+m-1)\}-u_0(i) 
    \end{equation}
    Where $u_0(i)$ is the average value and $i=\text{1,2,...,}N-m+1, m < N-2$ which is depicted in Eq. (\ref{eq:three_six}) \cite{three_nine}.
    \begin{equation} \label{eq:three_six}
        u_0(i)=\frac{1}{m}\sum_{j=0}^{m-1} u(i+j)
    \end{equation}
    \item Go through the Eq. (\ref{eq:five}) to (\ref{eq:nine}) as described in the standard FuEn (Eq. \ref{eq:nine}) \cite{three_nine}. 
\end{enumerate}
\subsubsection{Hierarchical Measure Entropy}
HFuEn is based on fuzzy entropy calculations and hierarchical procedure \cite{four_zero}. In this method, FuEn of each component is computed in the HFuEn analysis and then plotted as the function of the scale factor, which can be denoted as follows \cite{four_zero}.
\begin{equation}\label{eq:ur4}
    \text{HFuEn}(u,k,e,m,r) = FuEn(u_{k,e},m,r)
\end{equation}
In eq. (\ref{eq:ur4}), $u_{k,e}$ is the high frequency and low frequency component of time series $u(i)$ at scale $k$ and $k$ denotes kth layer of the hierarchical analysis \cite{four_zero}. More details about this HFuEn provided in reference \cite{four_zero}. 
\subsubsection{Fuzzy Measure Entropy}
This entropy is introduced by Chen et al. \cite{four_one}. At the research, FuMeEn is compared with ApEn, SampEn, and FuEn, and successful results are achieved. The FuMeEn is a generalization of FuEn and has a relation given by (\ref{eq:ur5}) \cite{four_one}:
\begin{equation}\label{eq:ur5}
\begin{split}
\text{FuzzyLMEN}(m,n_L,r_L,N) &= \\ 
\ln \phi L_m(n_L,r_L) &- \ln \phi L_{m+1}(n_L,r_L) \\
\text{FuzzyFMEN}(m,n_F,r_F,N) &= \\
\ln \phi L_m(n_F,r_F) &- \ln \phi L_{m+1}(n_F,r_F) 
\end{split}
\end{equation}
In the equations above, FuzzyLMEn and FuzzyFMEn are the calculations of Fuzzy Entropy with upper and lower boundaries, respectively. The final relation for entropy is defined as equation (\ref{eq:ur6}) \cite{four_one}:
\begin{equation}\label{eq:ur6}
    \text{FuzzyMEn}(m,n_L,r_L,n_F,r_F,N)=\text{FuzzyLMEN}+\text{FuzzyFMEN}
\end{equation}
The complete information on this entropy is given in \cite{four_one}.
\subsection{Computational Complexity of Features}
In this section, the computational complexity of these features are computed and compared. A standard method to calculate the computational complexity is using the big O notation (O) \cite{clrs}. While the hidden constants and underlying framework can dramatically affect algorithms' overall runtime, the big O gives the researchers enough insights to find the most suitable feature set for their work. 

As shown in \cite{two_seven}, FuEn can be calculated in $\text{O}(n^2m)$ where $m$ is the chosen length for patterns. For other advanced fuzzy entropies, usually, the base fuzzy entropy calculation is the dominant part, and their big O is similar to FuEn, i.e., $\text{O}(n^2m)$. However, their runtime could be different due to different hidden constants. Moreover, O for RCMFuE is of $\text{O}(\tau n^2m)$ and for MFuEn its the scaled of O(FuEn). Also, O for FuDistEn depends dramatically on the implementation of the density estimation calculator, and that part is usually the dominant part in O calculation.

\subsection{Dimensionality reduction using Autoencoder}
One of the aspects that can affect the performance of any classification model is the curse of dimensionality \cite{four_seven}. Many algorithms are suggested to reduce the dimensionality of the feature vector, from both categories of supervised learning, such as Fisher \cite{two_seven}, and unsupervised learning such as PCA \cite{four_eight,four_nine}. AEs \cite{four_two,four_three} are a group of unsupervised neural networks \cite{rotten2} that can be used for dimensionality reduction. In an AE, the input feature vector is first transformed into a smaller latent space with the encoder and then reconstructed from latent space with the decoder. Trained correctly, the encoder can be used as a dimensionality reducer. 

\begin{figure}[t]
    \centering
    \includegraphics[width=3.35in]{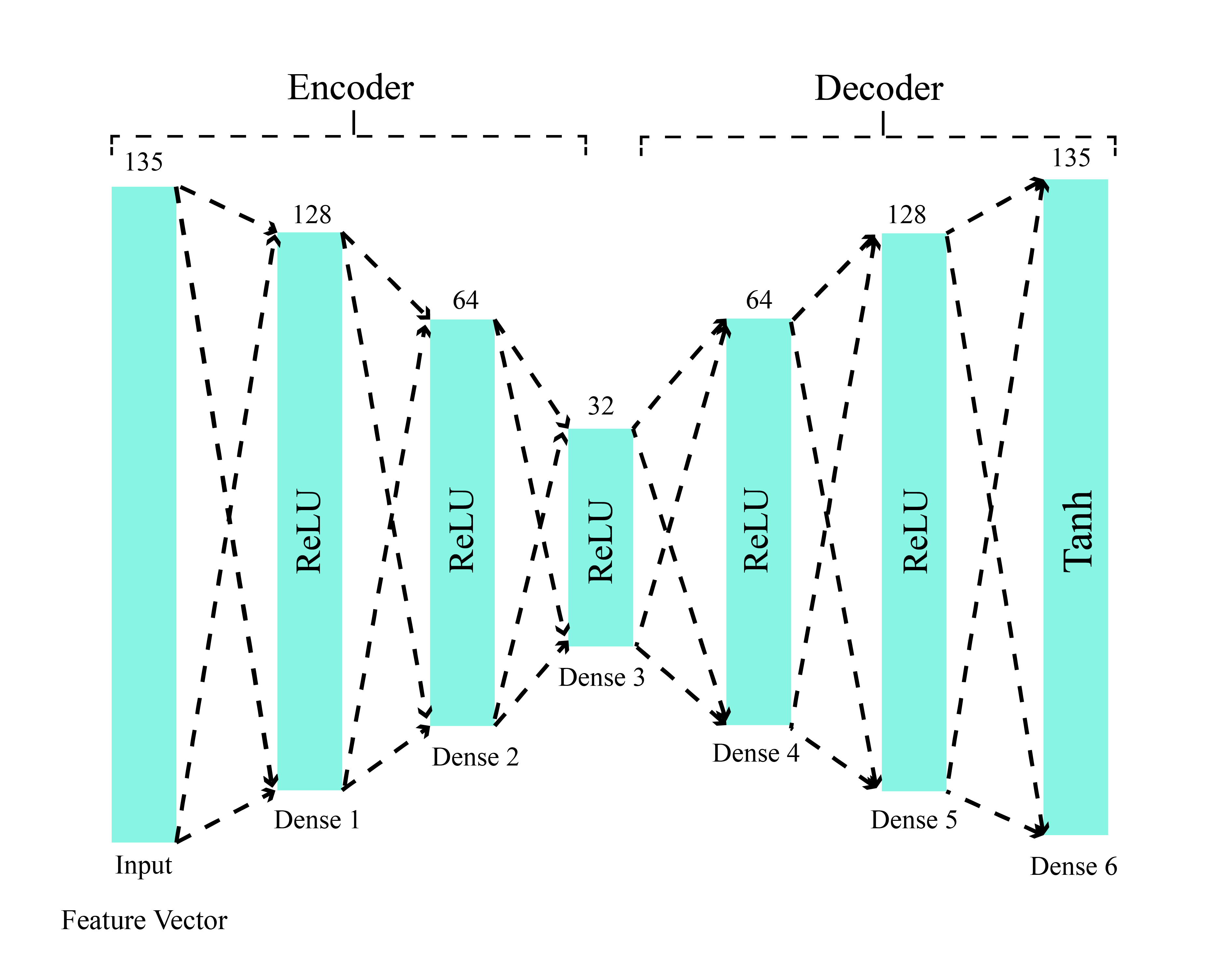}
    
    \caption{Proposed AE architecture for feature reduction.}
    \label{fig:four}
\end{figure}

To train this AE, an Adadelta \cite{five_zero} algorithm is applied as an optimizer and mean square error (MSE) as a loss function, respectively. For the last layer of AE, a tangent hyperbolic (Tanh) activation is used due to its faster convergence than the sigmoid activation function, as it has stronger gradients \cite{five_one}. Also, all data are normalized to match the output of the last layer activation (Tanh). The details of the AE network with the dimensionality reduction approach are given in Table \ref{tab:three}. In addition, Figure \ref{fig:four} shows the proposed AE architecture.

\begin{table}[t]
    \centering
    \caption{\textbf{Detailed parameter information of proposed AE}}
    \begin{tabular}{cccc}
        \Xhline{2\arrayrulewidth}
        \hline
        \hline
        \Xhline{2\arrayrulewidth}
        \thead{Layers} & \thead{Output Shape} & \thead{Parameters} & \thead{Activition}\\
        \Xhline{2\arrayrulewidth}
        Input & 135 & 0 & --- \\
        Dense-1 & 128 & 17408 & Relu \\
        Dense-2 & 64 & 8256 & Relu \\
        Dense-3 & 32 & 2080 & Relu \\
        Dense-4 & 64 & 2112 & Relu \\
        Dense-5 & 128 & 8320 & Relu \\
        Dense-6 & 135 & 17415 & Tanh\\
        \Xhline{2\arrayrulewidth}
        \hline
        \hline
        \Xhline{2\arrayrulewidth}
    \end{tabular}
    
    \label{tab:three}
\end{table}

\subsection{Classification}
\subsubsection{Standard ANFIS}

In recent decades, many different schemas have been used to design classification algorithms; such as statistical models \cite{astat}, SVMs \cite{asvm}, decision tree and its variations \cite{adt}, ensemble learning \cite{aw1}, fuzzy-based models \cite{afuz}, and ANNs \cite{amlp,aw2}. The result of combining neural network and fuzzy logic is a technique called ANFIS, which was proposed in 1993 \cite{four_four}. Fuzzy logic is very resembling human reasoning, and this significant issue has led to more advantages compared to neural networks. In ANFIS, analogously ANN, data inputs and labels can be applied to the network. In ANFIS, the system behavior is described by the membership function parameters from a dataset \cite{four_four}. A specified error condition is utilized for the adjustment of the parameters of the system. In the following, the structure and details of the ANFIS method are discussed.

\textbf{ANFIS Architecture} 

The learning and adaption are supported by the adaptive framework where ANFIS, Sugeno fuzzy strategy, is put into.\newline
A first-order Sugeno framework fuzzy model with the rules of IF-THEN is taken into account for the presentation of ANFIS design \cite{four_four}.
\newline

    \textbf{Rule 1:} if $x$ is $A_1$ and $y$ is $B_1$ then $f_1=p_1x+q_1y+r_1$
    
    \textbf{Rule 2:} if $x$ is $A_2$ and $y$ is $B_2$ then $f_2=p_2x+q_2y+r_2$
\newline

Where the inputs are $x$ and $y$, the fuzzy sets are $B_i$ and $A_i$. $p_i ,q_i$ and $r_i$ are the parameters of design that are denoted in training and $f_i$ are the outputs described by the fuzzy rule.

\begin{figure}[t]
    \centering
    \includegraphics[width=3.3in]{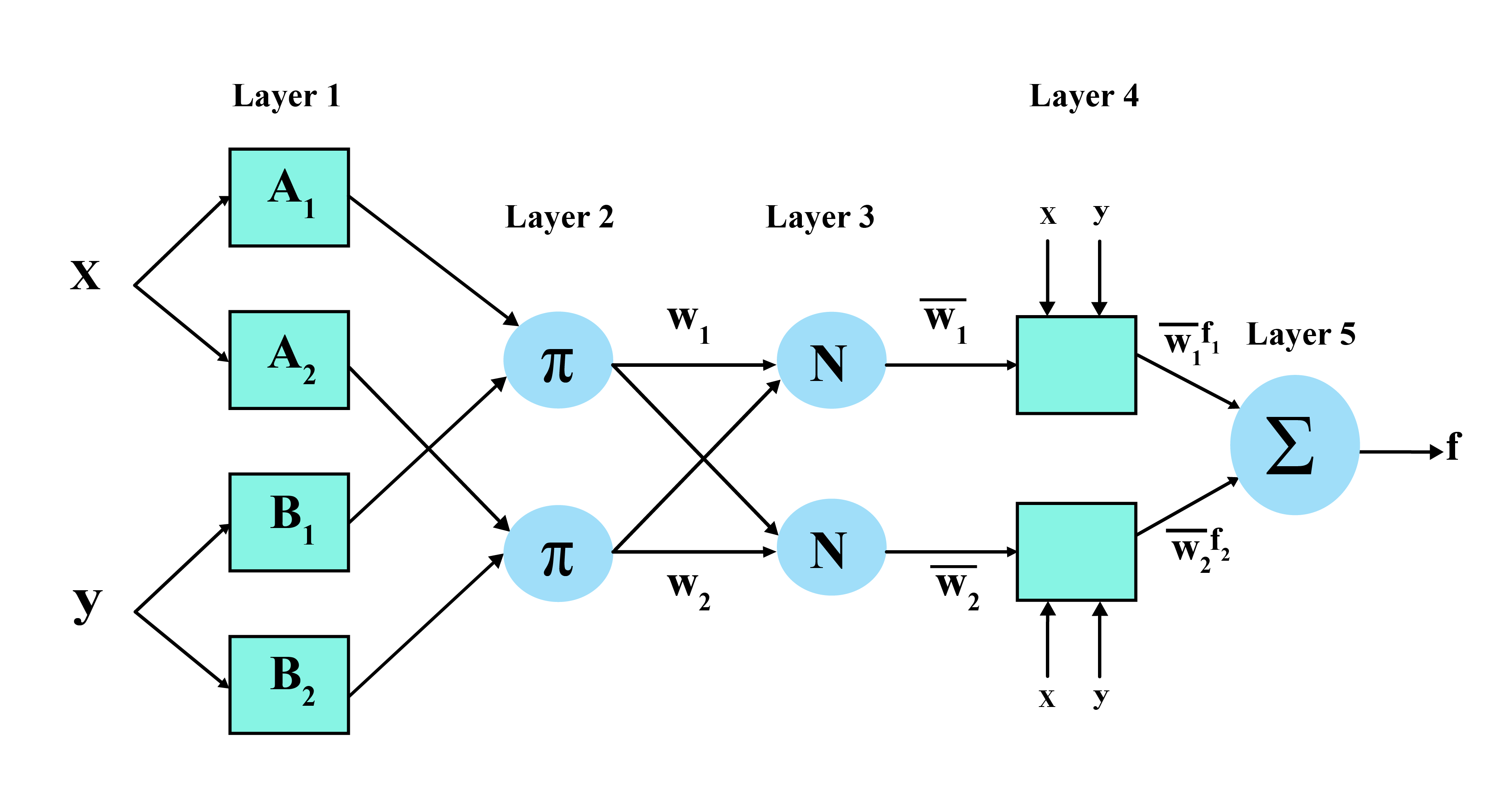}
    
    \caption{Block Diagram of standard ANFIS.}
    \label{fig:five}
\end{figure}

The individual layers are denoted as below with the assumption that an adaptive node is illustrated as square and a fixed node is drawn as a circle in Figure \ref{fig:five}. 

\textbf{Layer 1:} All the nodes $i$ in layer 1 are adaptive with a node function \cite{four_four}

\begin{equation} \label{eq:five_one}
    O_i^1=\mu_{A_i}(x)
\end{equation}

where $x$ is the input to node $i$, $\mu_{A_i}$  is the membership function of $A_i$. The membership function is selected as below \cite{four_four}:
\begin{equation} \label{eq:five_two}
    \mu_{A_i}(x)=\frac{1}{1+\left[ \left( \frac{x-c_i}{a_i} \right)^2 \right]^{b_i}}
\end{equation}

Or

\begin{equation}
    \mu_{A_i}(x)=\exp \left( - \left( \frac{x-c_i}{a_i} \right)^2 \right)
\end{equation}

where ${ a_i , b_i , c_i}$ is the principle parameter set and $x$ is the input \cite{four_four}.

\textbf{Layer 2:} The fixed nodes are presented in layer 2 and 3. In Layer 2, they operate as a simple multiplier and marked as M. The output is as follows for Layer 2 \cite{four_four}:
\begin{equation} \label{eq:five_three}
    O_i^2=\omega_i=\mu_{A_i}(x) . \mu_{B_i}(y), i=1,2
\end{equation}

These are the firing strength of the rules. 

\textbf{Layer 3:} The firing strength of layer 2 is normalized by the nodes and marked as N and represented as below \cite{four_four}:
\begin{equation}
    O_i^3=\overline{\omega}_i=\frac{\omega_i}{\omega_1+\omega_2}, i=1,2
\end{equation}

\textbf{Layer 4:} The product of a first-order polynomial and normalized firing strength is the output of Layer 4, where nodes are adaptive. The outputs are illustrated as below:
\begin{equation} \label{eq:five_five}
    O_i^4=\overline{\omega}_if_i=\overline{\omega}_i(p_ix+q_iy+r_i), i=1,2
\end{equation}

where $\overline{\omega}_i$ is the output of layer 3 and $\{p_i, q_i, r_i \}$ is the consequent parameter set \cite{four_four}.

\textbf{Layer 5:} There is a single fixed node marked as S in layer 5. All incoming signals are summed up by the node. The output as a whole is depicted as follows \cite{four_four}:
\begin{equation} \label{eq:five_six}
    O_i^5=\text{overalloutput}=\sum_i (\overline{\omega}_if_i)=\frac{\sum_i \omega_i f_i}{\sum_i \omega_i}
\end{equation}

\subsubsection{Optimal ANFIS Classifiers}

As mentioned earlier, fuzzy c-means clustering (FCM) is arguably the best method for conducting the fuzzy inference system (FIS) in ANFIS. Two schemas can be used to train ANFIS, namely, backpropagation and Hybrid methods \cite{four_four}. These methods directly affect membership functions, input, and outputs of the FIS in the training phase. Here, we used various optimization algorithms to improve ANFIS performance. In the suggested ANFIS, first, the membership functions' parameters are all put together in a vector; then, using PSO or BS optimizer, the best values are picked to minimize a chosen cost function. The cost function is described as follows:

\begin{equation}
    \min_{\theta} \text{Error} = \frac{1}{N}\sum_{i=1}^n e_i^2
\end{equation}
\begin{equation}
    e_i = t_i - f(x_i|\theta)
\end{equation}
\begin{equation}
    \text{RMSE} = \sqrt{\frac{1}{n} \sum_{i=1}^n (t_i - y_i)^2}
\end{equation}
Where $N$ is the number of ANFIS inputs, $e_i$ shows the error, $x_i$ shows input values, $\theta$ are ANFIS parameters, $n$ is the number of data instances, and $y_i$ is ANFIS output. Finally, GOA, PSO, and BS are used to minimize the error.

In this section, GOA \cite{goanew}, PSO \cite{five_two}, and BS \cite{five_three} optimization algorithms are applied separately for network training instead of back propagation procedure. The description of GOA and PSO algorithms can be found in their reference paper; also, the BS algorithm is explained below.
\begin{figure*}[t]
    \centering
    \includegraphics[width=6.5in]{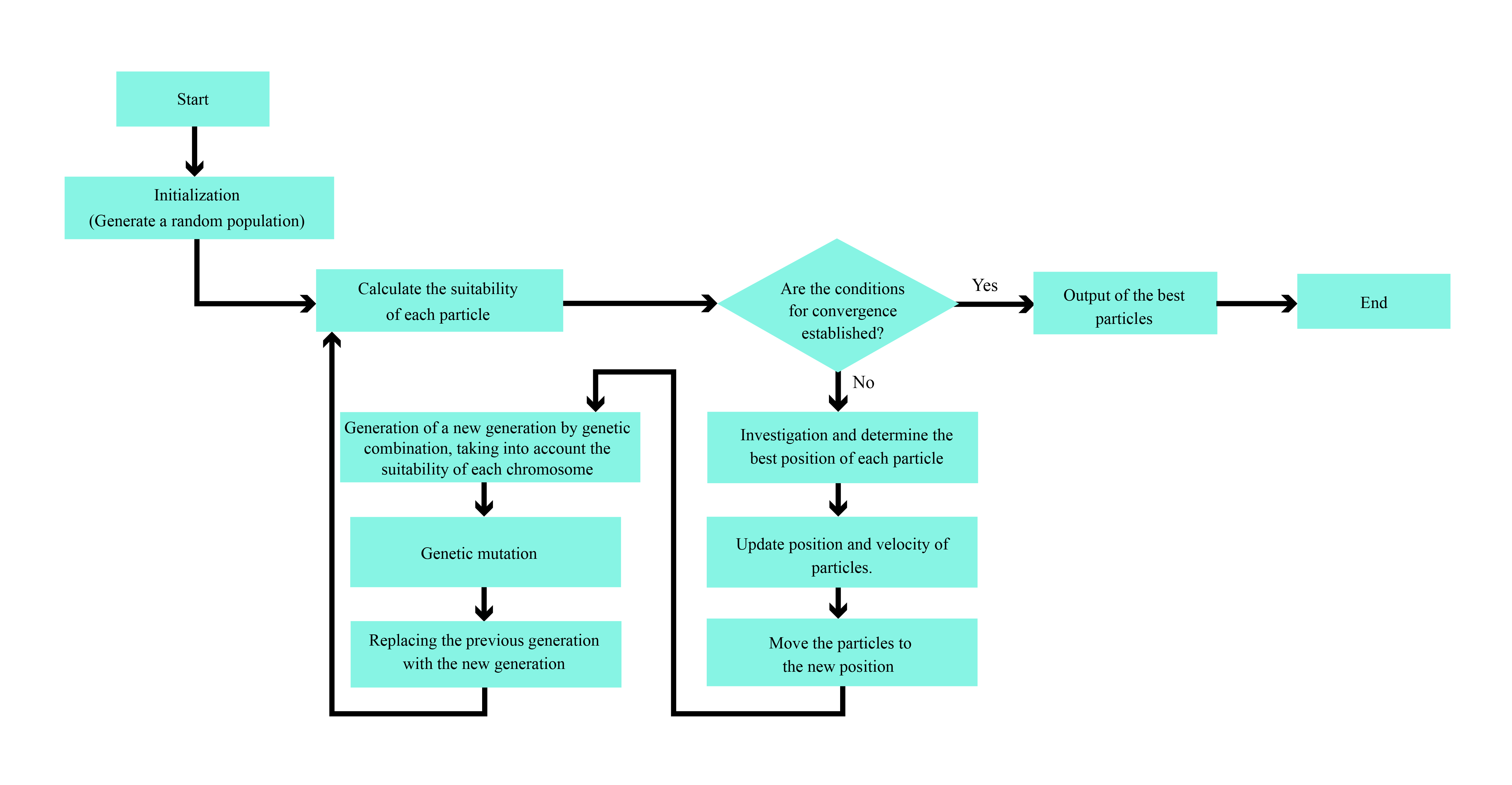}
    \caption{Flowchart of BS algorithm implementation.}
    \label{fig:seven}
\end{figure*}

\textbf{Breeding Swarm Optimization (BS)} 

In this section, the BS algorithm, which is a hybrid genetic algorithm (GA)-PSO algorithm, is presented. Both GA and PSO algorithms have advantages and drawbacks \cite{five_three}.

A PSO particle memorizes a part of the search space having better performance (a memory to store past experience), in GA, if a particle is not selected in the crossover or mutation phase, the relevant information of the particle is lost. Therefore, in the BS algorithm, the strengths of both PSO and GA algorithms are combined \cite{five_three}. The algorithm combined the steps of standard velocity and particle position update in PSO with the steps of selection, crossover, and genetic mutation in GA as a result, the GA section of the BS algorithm facilitates the global search, and the PSO section accomplishes the local search \cite{five_three}. The steps for implementing this algorithm are as follows \cite{five_three}:
\begin{itemize}
    \item \textbf{Step 1:} Generate a random population
    \item \textbf{Step 2:} Calculate the fitness of each particle according to the cost function
    \item \textbf{Step 3:} Select P best particles applying the roulette wheel algorithm
    \item \textbf{Step 4:} Perform step 3 of GA and PSO in parallel and creation a new population with PSO and GA outputs
    \item \textbf{Step 5:} Go to step 2 to achieve convergence.
\end{itemize}
Figure \ref{fig:seven} exhibits the block diagram of the BS algorithm steps.

\section{Statistical Parameters for Classification Performance}

The classification results are evaluated using 10-fold cross validation method. The advantage of K-fold cross-validation is that all the observations in the database are eventually used for both training and testing. Finally, the performance of the algorithm is estimated using evaluation metrics such as specificity (Spec), sensitivity (Sens), accuracy (Acc), and F1-Score (F-S) that is shown in Table \ref{tab:four}. These terminologies are extracted from the confusion matrix: true positive (TP), false negative (FN), true negative (TN), and false positive (FP) \cite{two_seven}.

\def\arraystretch{1.5}
\begin{table}[t]
    \centering
    \caption{\textbf{Description of performance parameters used}}
    \begin{tabular}{cc}
        \Xhline{2\arrayrulewidth}
        \hline
        \hline
        \Xhline{2\arrayrulewidth}
        \thead{Parameters Name} & \thead{Formula} \\
        \Xhline{2\arrayrulewidth}
        Accuracy & Acc = $\frac{\text{TP}+\text{TN}}{\text{FP}+\text{FN}+\text{TP}+\text{TN}} $ \\
        Sensivity & Sens = $\frac{\text{TP}}{\text{FP}+\text{TP}} $\\
        Specificity & Spec = $\frac{\text{TN}}{\text{FP}+\text{TN}} $\\
        Precision &  Prec = $\frac{\text{TP}}{\text{TP}+\text{FP}} $ \\
        F1\_score &  FS = $\frac{2\text{TP}}{2\text{TP}+\text{FP}+\text{FN}}$\\
        \Xhline{2\arrayrulewidth}
        \hline
        \hline
        \Xhline{2\arrayrulewidth}
    \end{tabular}
    
    \label{tab:four}
\end{table}
\def\arraystretch{1}

\begin{figure*}[t]
    \centering
    \includegraphics[width=7in]{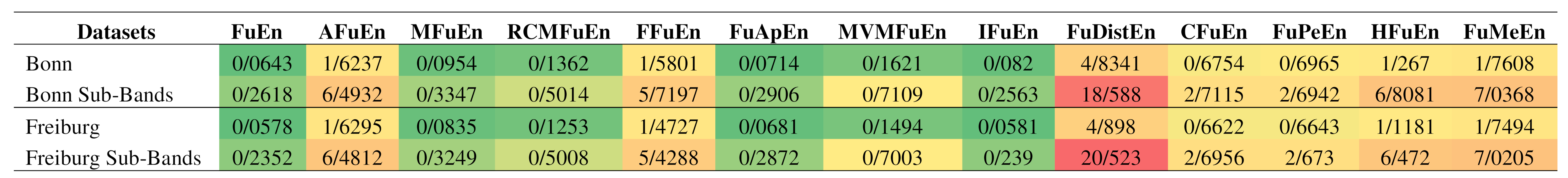}
    \caption{Runtime of different fuzzy entropies.}
    \label{fig:time}
\end{figure*}

\begin{table}[t]
    \centering
    \caption{\textbf{Hyper parameters used for optimization methods}}
    \begin{tabular}{ccc}
        \Xhline{2\arrayrulewidth}
        \hline
        \hline
        \Xhline{2\arrayrulewidth}
        \thead{PSO} & \thead{BS} & \thead{GOA} \\
        \Xhline{2\arrayrulewidth}
        $C1 = 2$ & $W1 = 1.8$ & $C_{\text{min}} = 0.00004$ \\
        $C2 = 2$ & $k1 = 2$ & $C_{\text{max}} = 1$\\
        $W = 0.2$ & $w = 0.2$\\
        \multicolumn{3}{c}{$\text{N\_{pop}}=60$} \\
        \multicolumn{3}{c}{$\text{Var\_min}=\text{ min(Feature\_Matrix)}$} \\
        \multicolumn{3}{c}{$\text{Var\_max} = \text{max(Feature\_Matrix)}$} \\
        \multicolumn{3}{c}{$\text{MAX\_{IT}}=400$} \\
        \Xhline{2\arrayrulewidth}
        \hline
        \hline
        \Xhline{2\arrayrulewidth}

    \end{tabular}
    \label{tab:five}
\end{table}

\section{Experiment Results}

In this section, the results of the proposed CADS based on fuzzy theory and deep learning techniques for the automated diagnosis of epileptic seizures are presented. The system used in this work had a Ryzen 1700 CPU, an Nvidia 1060 GPU, and 24GB of RAM. For software implementation of algorithms, Python 3.6 with Keras \cite{keras} is used for AE and Matlab 2019b for preprocessing, feature extraction, and classification. The identical preprocessing steps are adopted to preprocess the two datasets Bonn and Freiburg. First, the signals of both datasets are decomposed into the same time windows, and then TQWT is applied to decompose the signals into meaningful frequency sub-bands. At this stage, the windowing direction of the two datasets signals are selected in such a way as to achieve the highest level of efficiency and classification accuracy. In our previously conducted studies using the same database, a significant performance is achieved using EEG signals of the Bonn database with 5-seconds frame duration and outperformed other time frames \cite{eight,two_seven}. Hence, we have considered 5 seconds time window for this work also.  Additionally, in the Freiburg dataset, each signal is decomposed into 4-second time windows based on \cite{nb7,nb8,nb9}. As aforementioned in the previous sections, the TQWT has three important parameters $Q$, $r$, and $J$ to decompose the signals into different sub-bands. In this study, these parameters are selected as $Q = 1$, $r = 3$ and $J = 8$ for both datasets. In Figure \ref{fig:two}, a signal from the Bonn dataset using TQWT is decomposed and demonstrated.

In this study, the combination of various types of fuzzy entropies for feature extraction from EEG signals has been done for the first time. Each of the entropies involves a different computational complexity. To demonstrate the computational complexity of fuzzy entropies, they are employed on a frame of the Bonn and Freiburg datasets, and their execution times are shown in Figure \ref{fig:time}. According to Figure \ref{fig:time}, the minimum execution time is relevant to FuEn, and the highest time is allocated to FuDistEn. Figure \ref{fig:time} shows the execution time of each fuzzy entropy while applied to raw EEG signals and sub-bands of TQWT, respectively.

In this step, 15 fuzzy entropy were extracted from 9 sub-bands of TQWT. Therefore, for each frame of the EEG signals of the Bonn or Freiburg datasets, 135 features have been acquired. Next, in order to lessen the dimensions of the feature matrices of Bonn and Freiburg datasets, an AE network with the proposed number of layers has been used. Applying AE with dimensionality reduction allows the informative features to be selected and less effective features to be removed, which helps to enhance the performance of CADS. According to Figure \ref{fig:four} and Table \ref{tab:three}, after feature reduction, 32 features are obtained.

\begin{figure*}[t]
    \centering
    \includegraphics[width=6.5in]{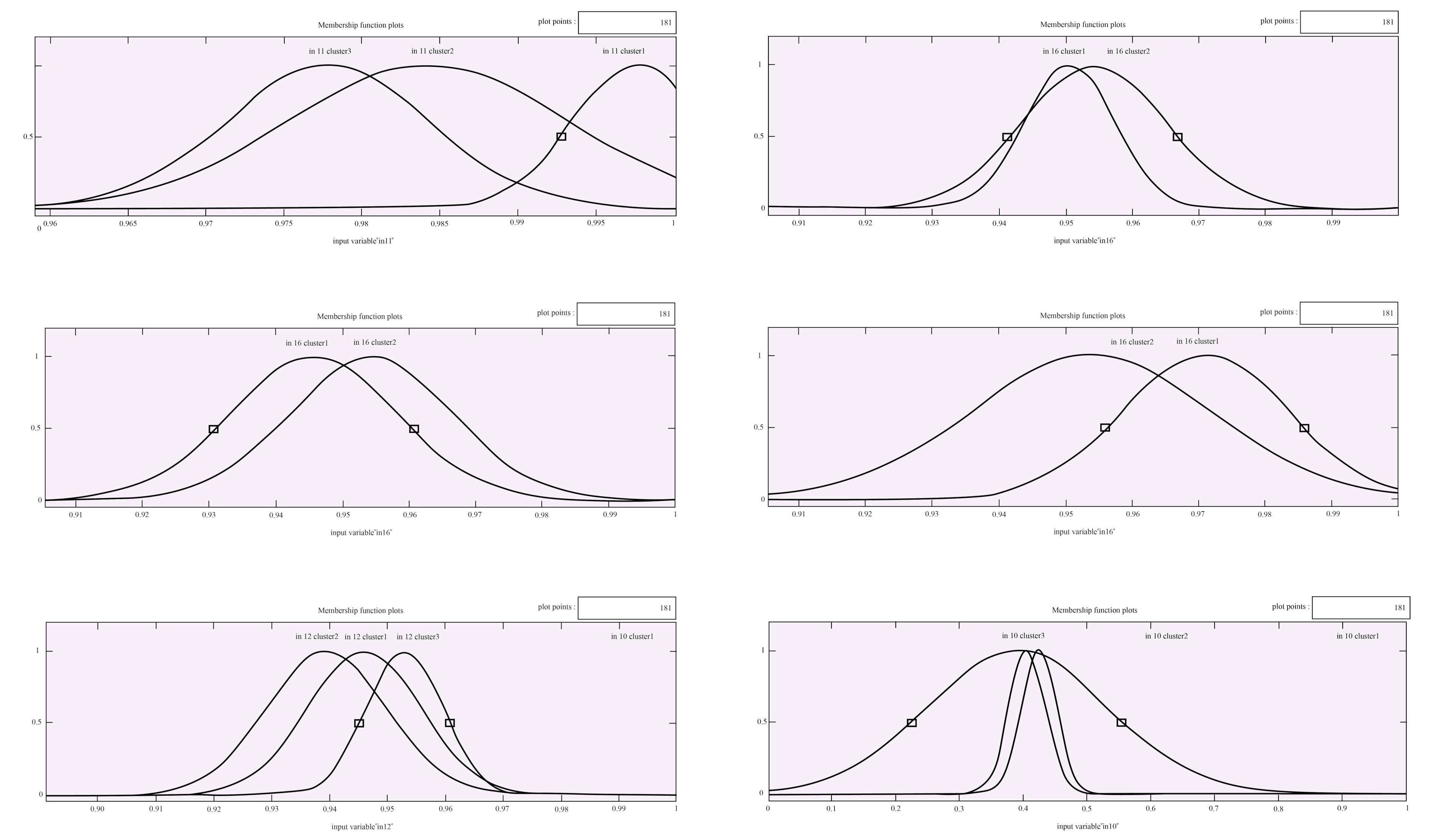}
    \caption{Samples of membership functions.}
    \label{fig:mem}
\end{figure*}

In the final step of the proposed CADS, a variety of different classifier algorithms were assayed containing two groups of fuzzy and non-fuzzy. Non-fuzzy methods embrace SVM, KNN, MLP, and RF. Also, fuzzy techniques for classification are based on the ANFIS method.
Takagi-Sugeno Kang (TSK) \cite{five_four} fuzzy system has been used to implement various classification procedures based on ANFIS. The basic fuzzy system of all these approaches is genfis-3 which is based on FCM method \cite{five_five,five_six}. The genfis-3 function is based on the Gaussian membership functions, and in our research, two and three membership functions are applied for each input. Figure \ref{fig:mem} displays a number of membership functions based on genfis-3 for some inputs of the Bonn and Freiburg datasets.

In this research, the standard ANFIS takes the advantage of various training procedures that the hybrid method has been exploited. As mentioned earlier, in ANFIS-PSO, ANFIS-GOA, and ANFIS-BS methods, an optimization method has been applied for the training. Therefore, the only difference between the standard ANFIS and its improved versions is in the training phase. Additionally, the important hyper parameters of PSO, GOA, and BS algorithms are given in Table \ref{tab:five}.

\begin{figure*}[t]
    \centering
    \includegraphics[width=6.5in]{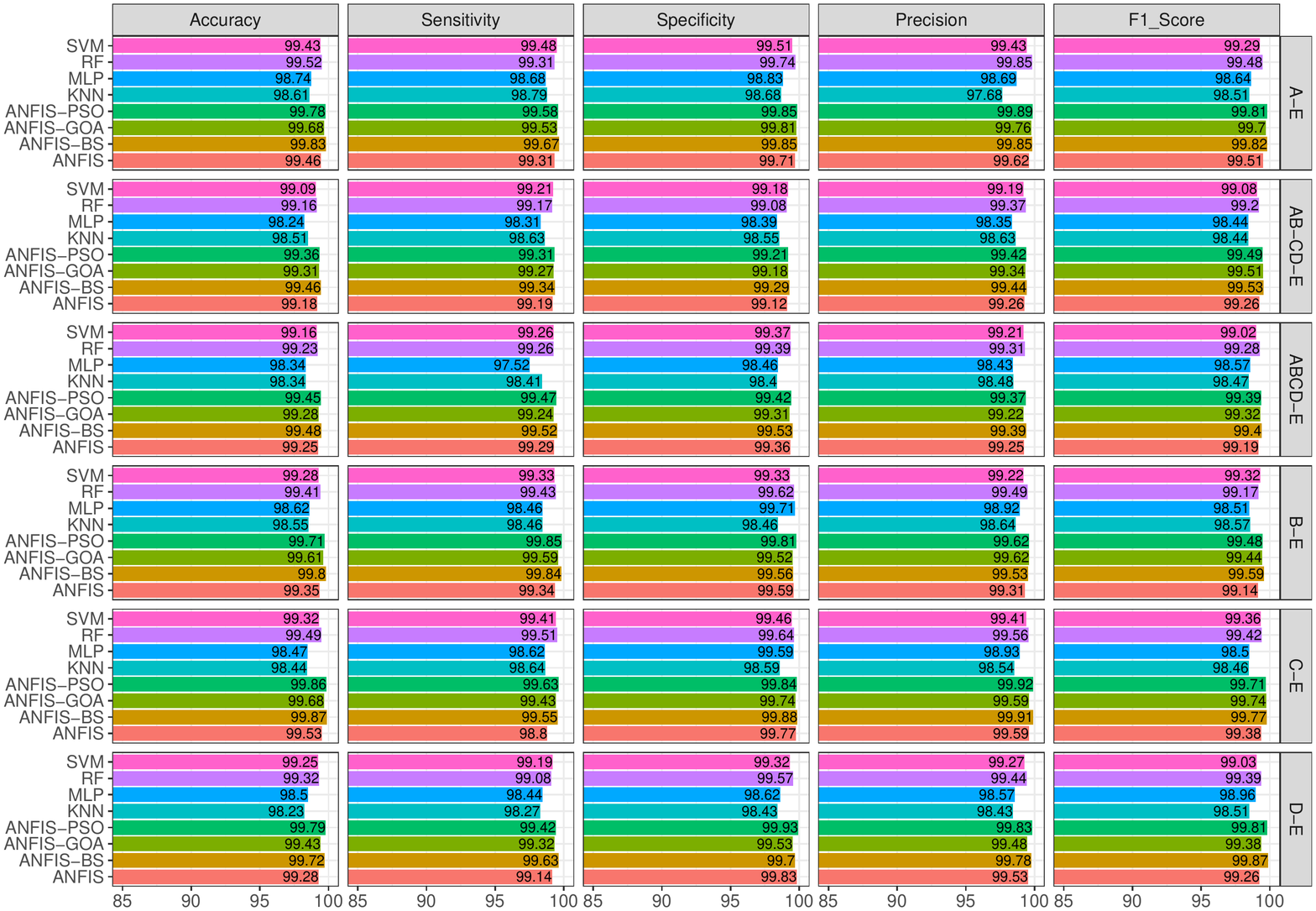}
    \caption{Summary of results obtained for various classifiers on Bonn dataset.}
    \label{fig:res}
\end{figure*}
\begin{table*}[ht]
    \centering
    \caption{\textbf{Summary of results obtained for various classifiers on Freiburg dataset}}
    \resizebox{0.83\linewidth }{!}{
    \begin{tabular}{cccccc}
    \Xhline{2\arrayrulewidth}
        \hline
        \hline
        \Xhline{2\arrayrulewidth}
        \thead{Method}  & \thead{Accuracy (\%)} & \thead{Sensitivity (\%)} & \thead{Specificity (\%)} & \thead{Precision (\%)} & \thead{F1\_Score (\%)} \\
        \Xhline{2\arrayrulewidth}
        SVM & 99.02 & 99.14 & 99.25 & 99.32 & 99.19 \\
        KNN & 98.19 & 98.66 & 98.53 & 98.20 & 98.34 \\
        MLP & 98.71 & 99.02 & 98.93 & 99.03 & 98.97 \\
        RF & 99.24 & 99.37 & 99.17 & 99.31 & 99.41 \\
        ANFIS & 99.16 & 99.44 & 99.36 & 99.19 & 99.26 \\
        ANFIS-PSO & 99.21 & 99.52 & 99.44 & 99.38 & 99.38 \\
        ANFIS-GOA & 99.19 & 99.41 & 99.48 & 99.13 & 99.31 \\
        ANFIS-BS & 99.28 & 99.54 & 99.56 & 99.29 & 99.49 \\
        \Xhline{2\arrayrulewidth}
        \hline
        \hline
        \Xhline{2\arrayrulewidth}
    \end{tabular}}
    
    \label{tab:fryres}
\end{table*}
In the following, the results of fuzzy and non-fuzzy classifier algorithms for each dataset are provided. In order to assess the performance of classification algorithms, each method is executed 10 times under identical conditions, and the evaluation parameters are reported on average. Carrying out this issue so makes the reported results more reliable and eliminates under/over performances due to chance. Figure \ref{fig:res} and Table \ref{tab:fryres} show the average results of different classifiers for the Bonn and Freiburg datasets, respectively. Results show that, the ANFIS-BS classifier achieved higher performances.

Applying the BS optimization method alongside ANFIS classification resulted in better performance of the ANFIS-BS method compared to others. In ANFIS, the backpropagation technique is used for training, which is based on gradient descent \cite{newer1}. Gradient descent moves toward the optima in the opposite direction of the loss function’s gradient. Choosing the correct learning rate dramatically affects the performance of this method; a large rate may cause the algorithm to overshoot the optima, while low rates may lead to small steps, thus the algorithm not learning anything \cite{newer2}. To solve this issue, researchers have used optimization techniques for the training of ANFIS and have reached promising results on many tasks \cite{newer3,newer4,newer5}; this paper is an example of these methods’ capabilities. As shown in Figure \ref{fig:res} and Table \ref{tab:fryres}, ANFIS-BS outperformed other classification methods. This is arguably due to the combination of GA and PSO in the BS algorithm; while BS algorithm has inherited the PSO speed benefits, also by adding GA, the performance has improved noticeably.

\begin{table*}[!h]
    \centering
    \caption{\textbf{Comparison of proposed method with other related works on Bonn dataset}}
    \resizebox{0.9\linewidth }{!}{
    \begin{tabular}{ccccc}
        \Xhline{2\arrayrulewidth}
        \hline
        \hline
        \Xhline{2\arrayrulewidth}
        \multirow{2}{*}{\thead{Works}} & \multirow{2}{*}{\thead{Methods}} & \multirow{2}{*}{\thead{K-fold}} & \multicolumn{2}{c}{\thead{Accuracy (\%)}} \\
        \cline{4-5}
        & & & \thead{Two Classes} & \thead{Multi Classes} \\
        \hline
 \cite{eight} & TQWT + FD, Entropy and Statistical Features + Ensemble Learning & 10 & 100 & 98 \\
 \cite{nine} & TQWT + CLP + RF & 10 & 99 & -- \\
 \cite{ten} & TQWT + QPS + NCA + KNN & 10 & 99.80 & 99.67 \\
%  \cite{one_one} & TQWT + Power Spectrum, Statistical, and Chaotic Features + Firefly Optimization + RF & 10 & 97 & -- \\
 \cite{one_two} & TQWT + Statistical Features + KNN & -- & 98.13 & -- \\
 \cite{one_thre} & TQWT + Multi-Scale KNN Entropy + SVM & 10 & 99.30 & 98.60 \\
%  \cite{one_four} & EMD – TQWT + IP Feature + LS-SVM & 10 & 99 & -- \\
 \cite{one_five} & TQWT + Spectral Features + Bagging & -- & 99 & 98.53 \\
 \cite{one_six} & TQWT + CCE Feature + Different Classifiers & -- & 98.30 & 98.20 \\
%  \cite{one_seven} & STFT + Spectrum Energy + Different Classifiers & 5 & 95.56 & -- \\
 \cite{one_eight} & DWT + Dynamic Features Based Entropy (Information entropy) + FSFS + LS-SVM & 10 & 99.50 & -- \\
 \cite{one_nine} & LSP + NCA Feature + SVM & 10 & 99.10 & 96.50 \\
 \cite{two_zero} & FBSE + WMRPE Feature + RF & 10 & 98.93 & -- \\
 \cite{two_one} & Matrix Determinant Based Features + MLP & 10 & 96.94 & 95.97 \\
 \cite{two_two} & IMFs + AmE + DESA + RF & 10 & 97.97 & 98.00 \\
 \cite{two_three} & DoG + LBP Feature + SVM & 10 & 99.41 & 98.80 \\
 \cite{two_four} & DWT + Different Features + RF & 10 & 98.75 & 91.39 \\
 \cite{two_five} & WPD + FuDistEn Feature + Kruskal-Wallis + KNN  & 10 & 99.69 & 99.07 \\
%  \cite{two_six} & GST + SVD Feature + RF & 10 & 99.12 & -- \\
 \cite{two_seven} & Filtering + Different Features + Fisher Score + CNN-AE & -- & 99.53 & 98.67 \\
\cite{a1} & Filtering, Normalization + 1D-CNN, Softmax & -- & 91.8 & 99 \\
\cite{a3} & Segmentation + DCT + Hurst Exponent and ARMA Model Features + LSTM, Softmax & -- & 97.78 & 94.81 \\
\cite{a4} & DWT + Feature Extraction + Stacked AE, Softmax & -- & 96 & --\\
\cite{a5} & Filtering + CNN-AE, Different Classifiers & 5/10 & 92 & --\\
\cite{a6} & Normalization + 1D-CNN, Softmax & 10 & -- & 98.67 \\
\cite{a7} & Filtering + RPS representation, AlexNet, Softmax & 10 & 98.5 & 95\\
% \cite{a8} & Normalization + CWT + mRMR Method + Pretrain Nets, KNN & 10 & 98.78 & --\\
\cite{a9} & Filtering + DWT + 2D-CNN, Softmax & -- & 97.74 & --\\
\cite{a10} & ApEn and RQA Features + 1D-CNN, Softmax & -- & 99.26 & --\\
\cite{a11} & Filtering + Combination of 1D-CNN and 2D-CNN, Softmax & 10 & 99.84 & 97.82\\
\cite{a12} & Normalization + Stacked Ensemble based DNN modeling, Meta Learner & 10 & 97.17 & --\\
\cite{new1} & Windowing + FuEn + K–S Two-Sample Test + SVM  & 10 & 100 & --\\
% \cite{new2} & Windowing + PSD, FuEn + SVM & -- & 96.2 & --\\
\cite{new3} & EMD + FuEn + SVM & -- & 97 & --\\
% \cite{new4} & Filtering, Windowing + FuPeEn + ANN & 10 & 98.72 & --\\
\cite{new5} & DWT + FuApEn + SVM & -- & 97.38 & --\\
% \cite{new6} & Windowing + mDistEn & 5 & 91 & --\\
% \cite{new7} & TQWT + mvFE + LS-SVM & 10 & 84.67 & --\\
\cite{new8} & FrFT-WPT + FuEn + PCA + SVM & 10 & 98.58& --\\
% \cite{new9} & Filtering, ICA, Windowing + MVMFuEn & -- & 100 & --\\
% \hline
% \cite{di1} & Butterworth filter + Fourier Transform + Wavelet Transform + T-test + SFFS + SVM & 5 & 100 & --\\
\cite{di2} & Filtering + Spike Detection Process + SVM, KNN, RF &  -- & 99.8 & -- \\
\cite{di3} & Filtering, DWT + Time Domain Features + NB, SVM & -- & 97.75 & --\\
\cite{di4} & \makecell{Filtering, HVD, DTCWT + Permutation Entropy, Spectral Entropy,\\Tsallis Entropy, Hjorth Parameters + relief Algorithm + M-SVM} & -- & 95.73 & --\\
 \cite{na1} & DWT + Statistical Features + ANN & -- & -- & 97.33\\
 \cite{na5} & TQWT + Correntropy Features + Kruskal Wallis, ANOVA + RF & -- & -- & 100\\
 \cite{na7} & Segmentation, Filtering, DWT, FFT + Statistical Features, PSD + ELM & -- & 99.68 & --\\
 \cite{na8} & Segmentation, WT + ApEn, LLE, Correlation Dimension + FRBS + LDAG-SVM & -- & -- & 95\\
 \cite{na10} & TsE + DT & -- & 100 & 92.67 \\
 \cite{na12} & \makecell{Segmentation, Clustering, Covariance Matrix + Statistical Features +\\Non-Parametric Tests + AB-LS-SVM} & -- & 99 & --\\
 \cite{na13} & CEEMD + MDE, RCMDE + Filter-Wrapper Based Method, One-Way-ANOVA + ANN & -- & 99 & 98.97\\
 \cite{na14} & Filtering, Segmentation, DWT + Temporal and Spectral Features + KNN, FRNN & -- & -- & --\\
 \cite{na17} & Filtering, VMD + Differential Entropies, PRMS + RF & -- & 94.1 & --\\
 \cite{na18} & Segmentation, CEEMD + Different Features + XGBoost & -- & 100 & 99\\
 \cite{na19} & Filtering, WD + FuEn, ReEn, KrEn + Kruskal-Wallis + Gaussian SVM & -- & 99.4 & --\\
 \cite{na20} & Filtering, Segmentation, DWT + Sigmoid Entropy + SVM & -- & 100  & --\\
 \cite{na21} & ECT + MGT, NPT, GLCM + PCA + RF & -- & 98 & --\\
%  \cite{na22} & Reshape, 2D-DWT + LGS + NCA + Different Classifiers & -- & -- & --\\
%  \cite{na26} & \makecell{Filtering, Segmentation, WVD, SVD + Histogram of\\Left Eigenvectors + Different Classifiers} & -- & -- & --\\
 \cite{na27} & Segmentation, WPD + Energy, ApEn + E-LPP + LS-SVM & -- & 99.5 & --\\
 \cite{na28} & Spectral Thresholding + Different Features + RF & -- & -- & 98.80\\
 \cite{na29} & DWT + Statistical Features + MEPA + APN & -- & 93.8 & --\\
 \cite{na30} & DWT + Statistical Features + CCP, PCA + LSTM & -- & 99 & --\\
 \cite{na31} & STFT + Haralick’s Texture Feature + DT & -- & 92.5 & --\\
 \makecell{Proposed \\ Method} & TQWT + Fuzzy Entropy Features Set + AE + ANFIS-BS & 10 & 99.74 & 99.46 \\
        \Xhline{2\arrayrulewidth}
        \hline
        \hline
        \Xhline{2\arrayrulewidth}
    \end{tabular}}
    \label{tab:seven}
\end{table*}

\begin{table*}[!h]
    \centering
    \caption{\textbf{Comparison of proposed method with other related works on Freiburg dataset}}
    \resizebox{0.83\linewidth }{!}{
    \begin{tabular}{ccc}
        \Xhline{2\arrayrulewidth}
        \hline
        \hline
        \Xhline{2\arrayrulewidth}
        \thead{Works}& \thead{Methods} &  \thead{Performance (\%)} \\
        \Xhline{2\arrayrulewidth}
        \cite{nb1}  & Filtering + ApEn, SampEn, PE, PFuzzy + SVM & Acc=95.3\\
        \cite{nb2}  & DWT + Energy, Entropy, STD, Mean + SVM & Acc=99.26\\
        \cite{nb3}  & DWT + Linear and Non-Linear Features + RF & Acc=97.74\\
        \cite{nb4}  & FFT + CNN & AUC=92\\
        \cite{nb5}  & Stockwell Transform + Bi-LSTM & Acc=98.69\\
        \cite{nb6} & DWT, DESA, Temporal and Spatial Averaging + FA + RF, Logistic, SVM & Acc=95\\
        \cite{nb7}  & Filtering + RL, FD, IV + Gradient Boosting & Sen=91.67\\
        \cite{nb8}  & DWT + R-ProCRC & --\\
        \cite{nb9}  & WPT + Relative Amplitude, PSD, PMRS + Weighted ELM & --\\
        \cite{nb10}  & Time and Frequency Domain Features + CNN & --\\
        \cite{nb11}  & Filtering, CBA + Linear and Non-Linear Features + SVM & Acc=96.8\\
        \cite{nb12}  & WT + Maximum, Minimum, Mean, STD + BoW + SVM\\
        \cite{nb13} & FT, WT + DCNN + Multi-View FCM & Acc=97.38\\
        \cite{nb14}  & DWT, S-Transform + CNN & Acc=98.12\\
        \cite{nb15} & Normalization, Filtering + LSTM & Acc=97.75\\
        \cite{nb16}  & Filtering + FC-NLSTM & Acc=96.17\\
        \cite{nb17}  & FFT, Filtering + Integer-Net & Acc=93.2\\
        \cite{nb18}  & \makecell{MSPCA, EMD, DWT, WPD + Statistical\\Features + RF, SVM, MLP, KNN} & Acc=100\\
        \cite{nb19}  & Filtering, Decomposition + FD, RFI + ELM-Trained SLFN & Acc=94.90\\
        \cite{nb20} & \makecell{Filtering, Normalization, Decomposition\\ + Different Features  + SVM} & Acc=97.5\\
        \cite{nb21}  & \makecell{Filtering, Normalization, HADTFD + TF-Flux,\\TF-Entropy, TF-Flatness + Spatial Averaging + Linear} & Acc=98.56\\
        \cite{nb22} & Filtering, LMD + RE, FI, Coefficient of Variation, + DPL & Acc=95.10 \\
        \cite{nb23} & DWT + Uniform 1D-LBP + SVM, ML-DF & Acc=95.33\\
        \cite{nb24} & Linear and Non-Linear Features + KH + GAN & Acc=98.9\\
        \cite{nb25} & MSPCA + PSD + DT & Acc=99.59\\
        \cite{nb26} & Different Methods + DPL & Acc=98.5\\
        \cite{nb27}  & Filtering + Different Features + ANOVA, Tukey’s Post-Hoc & --\\
         \makecell{Proposed \\ Method} & TQWT + Fuzzy Entropy Features Set + AE + ANFIS-BS & Acc=99.28 \\
        \Xhline{2\arrayrulewidth}
        \hline
        \hline
        \Xhline{2\arrayrulewidth}
    \end{tabular}}
    \label{tab:fryrel}
\end{table*}
\section{Limitations of the Study}

This section is devoted to discussing the limitations of the proposed method. For the first limitation, the high computational cost in the feature extraction can limit the applications of the proposed system; as a solution to this problem, the method can be implemented on hardware such as field programmable gate array, with appropriate approximations to increase the speed. The second limitation relates to setting hyper-parameters in fuzzy entropies; given that nonlinear feature extraction methods such as fuzzy entropies require precise parameter adjustments to achieve high accuracy results, they are usually chosen by trial and error, limiting their performance to the accuracy of these trials. Other limitations of the proposed method are the use of ANFIS for classification applications. 

As mentioned in the previous sections, the ANFIS classification method's optimization has been done using GOA, PSO, and BS algorithms. Precisely adjusting the parameters of PSO and BS algorithms to increase the performance of ANFIS is time-consuming and requires sufficient knowledge in the field of optimization and fuzzy algorithms. In this paper, ANFIS, ANFIS-GOA, ANFIS-PSO, and ANFIS-BS classification techniques are implemented based on the TSK model. In MATLAB, the basic TSK system is implemented based on the Gaussian membership function, which is a limitation, considering that other types of membership functions such as triangular and trapezoidal can not be examined and tested in the proposed classification methods.

\section{Discussion, Conclusion and Future Directions}

Diagnosis of epileptic seizures in the early stages is of particular significance to physicians and neurologists. Early detection of epileptic seizures allows specialist physicians to control the disease and prevent further progression.

So far, various methods have been proposed for epileptic seizures detection, among which the recording of EEG signals has received a great deal of attention among physicians and neurologists. EEG signals provide physicians with significant information about the function of brain activity. The most important advantages of EEG recording possess low cost and accurate display of brain neuronal function for accurate diagnosis of epileptic seizures. However, diagnosing epileptic seizures based on EEG signals is always a challenging task for physicians. The high complexity, the presence of different noises, and the long-term EEG signal recording are among the most important problems that make it difficult for specialist physicians or neurologists to precisely and quickly diagnose epileptic seizures \cite{rrt5}.

In recent years, researchers have tried using AI techniques coupled with EEG signals to assist physicians in diagnosing epileptic seizures faster and more accurately. So far, much research has been conducted on the implementation of CAD systems based on AI for epileptic seizure diagnosis \cite{rrt1,rrt2,rrt3,rrt4,rrt5}. AI-based CAD systems for diagnosing epileptic seizures include both conventional machine learning and deep learning \cite{rrt1,rtltrt1}.

In this paper, a novel CADS is introduced for epileptic seizures detection from EEG signals using fuzzy theory and deep learning methods. The proposed method involves four steps of preprocessing, feature extraction, dimensionality reduction, and classification. Preprocessing consists of two parts: windowing the EEG signals and then decomposing them into different sub-bands using TQWT. First, the EEG signals of the Bonn and Freiburg datasets are decomposed into various time windows. The objective is to achieve maximum CADS performance for two different datasets. The EEG signals of the Bonn dataset are segmented into 5-seconds time windows similar to \cite{eight,two_seven}. Additionally, the EEG signals of the Freiburg dataset have been decomposed into 4-second time windows based on \cite{nb7,nb8,nb9}. In the second stage of preprocessing, TQWT is exploited to decompose the EEG signals of both datasets into 9 sub-bands. EEG signals possess fundamental information in different frequency bands. Consequently, the decomposition of EEG signals into various sub-bands using TQWT enhances the efficiency of CADS in epileptic seizures detection. TQWT has three main parameters, $r$, $J$, and $Q$; selected values for both datasets are $r = 3$, $Q = 1$, and $J = 8$.

In the next step, type-1 fuzzy entropies are applied for feature extraction from different sub-bands of TQWT. In all previous works \cite{three,five_seven,five_eight,four_six}, this combination of fuzzy entropy types is not used for feature extraction, and this step is one of the novelties of our paper. Also, in this stage, the computational complexity for each fuzzy entropy is calculated as another novelty of this work. Calculating the computational time complexity of feature extraction algorithms is very important for signal processing researchers. So far, in several studies, computational complexity in feature extraction algorithms has been considered by researchers \cite{rro1,rro2,rro3,rro4,rro5}. In this research, for the first time, the computational time complexity for fuzzy entropies is presented. In addition, Figure \ref{fig:time} displays the run times for each fuzzy entropy, which is highly important for practical applications and hardware implementations. The proposed AE architecture with 6-layers is presented in the dimensionality reduction section, which is another novelty of this work.

Finally, various fuzzy and non-fuzzy techniques have been applied in the classification step. In this section, the ANFIS-GOA and ANFIS-BS methods have been employed for the first time in epileptic seizures detection and are another novelty. The ANFIS-BS method outperforms ANFIS-PSO and ANFIS-GOA. This verifies that the choice of an optimization method along with ANFIS demands high knowledge in the fuzzy systems theory and optimization methods.

The proposed method results show that the ANFIS-BS classifier method has been able to reach the highest accuracy among all fuzzy and non-fuzzy classifiers for its two different datasets. The reason behind the high efficiency of the ANFIS-BS procedure against all fuzzy and non-fuzzy classifier algorithms is that the BS optimization algorithm is a combination of GA and PSO methods. In the Bonn dataset, the best performance is acquired using ANFIS-BS, which in the two-class and three-class mode are 99.74\% and 99.46\%, respectively. Also, the ANFIS-BS method on the Freiburg dataset provides 99.28\% accuracy, which is the highest efficiency compared to other classifier methods. In addition, it should be reminded that in this study, we have used the K-fold method with k equal to 10 for evaluation in all classifier methods.

In Tables \ref{tab:seven} and \ref{tab:fryrel}, the research papers conducted on Bonn and Freiburg datasets for epileptic seizures detection based on AI techniques are presented and compared with the proposed method. This section provides a complete overview of papers on the Bonn and Freiburg datasets.

According to Tables \ref{tab:seven} and \ref{tab:fryrel}, our proposed method has achieved higher accuracy than most studies. The results indicate that the proposed technique compared to other studies, has attained an acceptable accuracy in diagnosing epileptic seizures. The proposed approach takes advantage of a combination of fuzzy logic and deep learning and has been successful in diagnosing epileptic seizures. The proposed method can aid physicians and other relevant medical professionals in the future as a software or hardware platform in diagnosing and predicting epileptic seizures using EEG signals. It can be help physicians in the rapid and accurate diagnosis of epileptic seizures in the near future, which will reduce the time to diagnose epileptic seizures.

For future directions, investigating the performance of the method on more complicated datasets or clinical ones or employing the method on more sophisticated tasks, such as seizure prediction, can help to examine the potential of the technique more accurately. Designing robotic systems and applications for smartphones to help patients with epilepsy, creating real-time tools to help experts in more accurate and faster detection of seizures, and combining deep learning methods, such as convolutional neural networks (CNN) with fuzzy-based features for feature fusion, feeding extracted features to long short term memory (LSTM) models or employing a hierarchical structure consisting of both deep learning and fuzzy entropies to minimize uncertainty \cite{rtltrt2}, all can be investigated as future directions to further improve the performance.

Additionally, in future works, type-2 fuzzy methods can replace the ANFIS \cite{newer6,newer7,newer8}. Nevertheless, employing type-1 and type-2 fuzzy regression algorithms to predict seizures is another direction for future studies \cite{newer9,newer10,newer11}.

\bibliographystyle{IEEEtran}
\bibliography{main}

\begin{thebibliography}{100}
\providecommand{\url}[1]{#1}
\csname url@samestyle\endcsname
\providecommand{\newblock}{\relax}
\providecommand{\bibinfo}[2]{#2}
\providecommand{\BIBentrySTDinterwordspacing}{\spaceskip=0pt\relax}
\providecommand{\BIBentryALTinterwordstretchfactor}{4}
\providecommand{\BIBentryALTinterwordspacing}{\spaceskip=\fontdimen2\font plus
\BIBentryALTinterwordstretchfactor\fontdimen3\font minus
  \fontdimen4\font\relax}
\providecommand{\BIBforeignlanguage}[2]{{%
\expandafter\ifx\csname l@#1\endcsname\relax
\typeout{** WARNING: IEEEtran.bst: No hyphenation pattern has been}%
\typeout{** loaded for the language `#1'. Using the pattern for}%
\typeout{** the default language instead.}%
\else
\language=\csname l@#1\endcsname
\fi
#2}}
\providecommand{\BIBdecl}{\relax}
\BIBdecl

\bibitem{one}
S.~M. Qaisar and A.~Subasi, ``Effective epileptic seizure detection based on
  the event-driven processing and machine learning for mobile healthcare,''
  \emph{Journal of Ambient Intelligence and Humanized Computing}, pp. 1--13,
  2020.

\bibitem{oner}
T.~Tuncer, S.~Dogan, F.~Ertam, and A.~Subasi, ``A novel ensemble local graph
  structure based feature extraction network for eeg signal analysis,''
  \emph{Biomedical Signal Processing and Control}, vol.~61, p. 102006, 2020.

\bibitem{added1}
A.~R. Hassan, A.~Subasi, and Y.~Zhang, ``Epilepsy seizure detection using
  complete ensemble empirical mode decomposition with adaptive noise,''
  \emph{Knowledge-Based Systems}, vol. 191, p. 105333, 2020.

\bibitem{addedr1}
A.~Subasi, J.~Kevric, and M.~A. Canbaz, ``Epileptic seizure detection using
  hybrid machine learning methods,'' \emph{Neural Computing and Applications},
  vol.~31, no.~1, pp. 317--325, 2019.

\bibitem{ttwo_ei}
E.~Alickovic, J.~Kevric, and A.~Subasi, ``Performance evaluation of empirical
  mode decomposition, discrete wavelet transform, and wavelet packed
  decomposition for automated epileptic seizure detection and prediction,''
  \emph{Biomedical signal processing and control}, vol.~39, pp. 94--102, 2018.

\bibitem{ttwo_eier}
A.~R. Hassan and A.~Subasi, ``Automatic identification of epileptic seizures
  from eeg signals using linear programming boosting,'' \emph{computer methods
  and programs in biomedicine}, vol. 136, pp. 65--77, 2016.

\bibitem{riper1}
A.~Subasi, ``Application of adaptive neuro-fuzzy inference system for epileptic
  seizure detection using wavelet feature extraction,'' \emph{Computers in
  biology and medicine}, vol.~37, no.~2, pp. 227--244, 2007.

\bibitem{ntwo2}
L.~A. Allen, R.~M. Harper, S.~Lhatoo, L.~Lemieux, and B.~Diehl, ``Neuroimaging
  of sudden unexpected death in epilepsy (sudep): Insights from structural and
  resting-state functional mri studies,'' \emph{Frontiers in neurology},
  vol.~10, p. 185, 2019.

\bibitem{ntwo3}
K.~Blackmon, ``Structural mri biomarkers of shared pathogenesis in autism
  spectrum disorder and epilepsy,'' \emph{Epilepsy \& Behavior}, vol.~47, pp.
  172--182, 2015.

\bibitem{two}
U.~R. Acharya, H.~Fujita, V.~K. Sudarshan, S.~Bhat, and J.~E. Koh,
  ``Application of entropies for automated diagnosis of epilepsy using eeg
  signals: A review,'' \emph{Knowledge-Based Systems}, vol.~88, pp. 85--96, 11
  2015.

\bibitem{three}
A.~Shoeibi, N.~Ghassemi, M.~Khodatars, M.~Jafari, S.~Hussain, R.~Alizadehsani,
  P.~Moridian, A.~Khosravi, H.~Hosseini-Nejad, M.~Rouhani \emph{et~al.},
  ``Epileptic seizure detection using deep learning techniques: A review,''
  \emph{arXiv preprint arXiv:2007.01276}, 2020.

\bibitem{four}
M.~Khodatars, A.~Shoeibi, N.~Ghassemi, M.~Jafari, A.~Khadem, D.~Sadeghi,
  P.~Moridian, S.~Hussain, R.~Alizadehsani, A.~Zare \emph{et~al.}, ``Deep
  learning for neuroimaging-based diagnosis and rehabilitation of autism
  spectrum disorder: A review,'' \emph{arXiv preprint arXiv:2007.01285}, 2020.

\bibitem{five}
M.~K. Islam, A.~Rastegarnia, and Z.~Yang, ``A wavelet-based artifact reduction
  from scalp eeg for epileptic seizure detection,'' \emph{IEEE Journal of
  Biomedical and Health Informatics}, vol.~20, pp. 1321--1332, 09 2016.

\bibitem{six}
N.~Kannathal, M.~L. Choo, U.~R. Acharya, and P.~Sadasivan, ``Entropies for
  detection of epilepsy in eeg,'' \emph{Computer Methods and Programs in
  Biomedicine}, vol.~80, pp. 187--194, 12 2005.

\bibitem{seven}
\BIBentryALTinterwordspacing
U.~R. Acharya, Y.~Hagiwara, and H.~Adeli, ``Automated seizure prediction,''
  \emph{Epilepsy \& Behavior}, vol.~88, pp. 251--261, 11 2018. [Online].
  Available:
  \url{https://www.sciencedirect.com/science/article/pii/S1525505018305791}
\BIBentrySTDinterwordspacing

\bibitem{two_eight}
I.~W. Selesnick, ``Wavelet transform with tunable q-factor,'' \emph{IEEE
  transactions on signal processing}, vol.~59, no.~8, pp. 3560--3575, 2011.

\bibitem{two_nine}
\BIBentryALTinterwordspacing
Forscher, ``Epileptologie bonn / forschung / ag lehnertz / eeg data download,''
  www.meb.uni-bonn.de. [Online]. Available:
  \url{http://www.meb.uni-bonn.de/epileptologie/science/physik/eegdata.html.}
\BIBentrySTDinterwordspacing

\bibitem{nb1}
H.~Waqar, J.~Xiang, M.~Zhou, T.~Hu, B.~Ahmed, S.~H. Shapor, M.~S. Iqbal, and
  M.~Raheel, ``Towards classifying epileptic seizures using entropy variants,''
  in \emph{2019 IEEE Fifth International Conference on Big Data Computing
  Service and Applications (BigDataService)}.\hskip 1em plus 0.5em minus
  0.4em\relax IEEE, 2019, pp. 296--300.

\bibitem{eight}
N.~Ghassemi, A.~Shoeibi, M.~Rouhani, and H.~Hosseini-Nejad, ``Epileptic
  seizures detection in eeg signals using tqwt and ensemble learning,''
  \emph{2019 9th International Conference on Computer and Knowledge Engineering
  (ICCKE)}, 10 2019.

\bibitem{ent}
H.~Ocak, ``Automatic detection of epileptic seizures in eeg using discrete
  wavelet transform and approximate entropy,'' \emph{Expert Systems with
  Applications}, vol.~36, no.~2, pp. 2027--2036, 2009.

\bibitem{fractal}
G.~E. Polychronaki, P.~Y. Ktonas, S.~Gatzonis, A.~Siatouni, P.~Asvestas,
  H.~Tsekou, D.~Sakas, and K.~Nikita, ``Comparison of fractal dimension
  estimation algorithms for epileptic seizure onset detection,'' \emph{Journal
  of neural engineering}, vol.~7, no.~4, p. 046007, 2010.

\bibitem{cor}
J.~R. Williamson, D.~W. Bliss, D.~W. Browne, and J.~T. Narayanan, ``Seizure
  prediction using eeg spatiotemporal correlation structure,'' \emph{Epilepsy
  \& behavior}, vol.~25, no.~2, pp. 230--238, 2012.

\bibitem{llp}
S.~Osowski, B.~Swiderski, A.~Cichocki, and A.~Rysz, ``Epileptic seizure
  characterization by lyapunov exponent of eeg signal,'' \emph{COMPEL-The
  international journal for computation and mathematics in electrical and
  electronic engineering}, 2007.

\bibitem{new1}
J.~Xiang, C.~Li, H.~Li, R.~Cao, B.~Wang, X.~Han, and J.~Chen, ``The detection
  of epileptic seizure signals based on fuzzy entropy,'' \emph{Journal of
  neuroscience methods}, vol. 243, pp. 18--25, 2015.

\bibitem{new2}
G.~Srivastava, A.~Tripathi, and P.~Maurya, ``Fuzzy entropy based seizure
  detection algorithms for eeg data analysis,'' in \emph{Smart Healthcare for
  Disease Diagnosis and Prevention}.\hskip 1em plus 0.5em minus 0.4em\relax
  Elsevier, 2020, pp. 89--101.

\bibitem{new3}
D.~Tripathi and N.~Agrawal, ``Epileptic seizure detection using empirical mode
  decomposition based fuzzy entropy and support vector machine,'' in
  \emph{International Conference on Green and Human Information
  Technology}.\hskip 1em plus 0.5em minus 0.4em\relax Springer, 2018, pp.
  109--118.

\bibitem{two_five}
T.~Zhang, W.~Chen, and M.~Li, ``Fuzzy distribution entropy and its application
  in automated seizure detection technique,'' \emph{Biomedical Signal
  Processing and Control}, vol.~39, pp. 360--377, 01 2018.

\bibitem{new4}
W.~Hussain, B.~Wang, Y.~Niu, Y.~Gao, X.~Wang, J.~Sun, Q.~Zhan, R.~Cao,
  Z.~Mengni, M.~S. Iqbal \emph{et~al.}, ``Epileptic seizure detection with
  permutation fuzzy entropy using robust machine learning techniques,''
  \emph{IEEE Access}, vol.~7, pp. 182\,238--182\,258, 2019.

\bibitem{three_nine}
W.~Hussain, B.~Wang, Y.~Niu, Y.~Gao, X.~Wang, J.~Sun, Q.~Zhan, R.~Cao,
  Z.~Mengni, M.~S. Iqbal \emph{et~al.}, ``Epileptic seizure detection with
  permutation fuzzy entropy using robust machine learning techniques,''
  \emph{IEEE Access}, vol.~7, pp. 182\,238--182\,258, 2019.

\bibitem{new5}
Y.~Kumar, M.~Dewal, and R.~Anand, ``Epileptic seizure detection using dwt based
  fuzzy approximate entropy and support vector machine,''
  \emph{Neurocomputing}, vol. 133, pp. 271--279, 2014.

\bibitem{new7}
A.~Bhattacharyya, R.~B. Pachori, and U.~R. Acharya, ``Tunable-q wavelet
  transform based multivariate sub-band fuzzy entropy with application to focal
  eeg signal analysis,'' \emph{Entropy}, vol.~19, no.~3, p.~99, 2017.

\bibitem{new8}
M.~Li, W.~Chen, and T.~Zhang, ``Fuzzyen-based features in frft-wpt domain for
  epileptic seizure detection,'' \emph{Neural Computing and Applications},
  vol.~31, no.~12, pp. 9335--9348, 2019.

\bibitem{three_zero}
W.~Chen, J.~Zhuang, W.~Yu, and Z.~Wang, ``Measuring complexity using fuzzyen,
  apen, and sampen,'' \emph{Medical engineering \& physics}, vol.~31, no.~1,
  pp. 61--68, 2009.

\bibitem{three_one}
J.-M. Girault and A.~Humeau-Heurtier, ``Centered and averaged fuzzy entropy to
  improve fuzzy entropy precision,'' \emph{Entropy}, vol.~20, no.~4, p. 287,
  2018.

\bibitem{three_two}
J.~Zheng, H.~Pan, and J.~Cheng, ``Rolling bearing fault detection and diagnosis
  based on composite multiscale fuzzy entropy and ensemble support vector
  machines,'' \emph{Mechanical Systems and Signal Processing}, vol.~85, pp.
  746--759, 2017.

\bibitem{three_three}
H.~Azami and J.~Escudero, ``Refined composite multivariate generalized
  multiscale fuzzy entropy: A tool for complexity analysis of multichannel
  signals,'' \emph{Physica A: Statistical Mechanics and its Applications}, vol.
  465, pp. 261--276, 2017.

\bibitem{three_four}
S.~He, K.~Sun, and R.~Wang, ``Fractional fuzzy entropy algorithm and the
  complexity analysis for nonlinear time series,'' \emph{The European Physical
  Journal Special Topics}, vol. 227, no. 7-9, pp. 943--957, 2018.

\bibitem{three_five}
Y.~Kumar, M.~Dewal, and R.~Anand, ``Epileptic seizure detection using dwt based
  fuzzy approximate entropy and support vector machine,''
  \emph{Neurocomputing}, vol. 133, pp. 271--279, 2014.

\bibitem{three_six}
S.~Raghu, N.~Sriraam, G.~P. Kumar, and A.~S. Hegde, ``A novel approach for
  real-time recognition of epileptic seizures using minimum variance modified
  fuzzy entropy,'' \emph{IEEE Transactions on Biomedical Engineering}, vol.~65,
  no.~11, pp. 2612--2621, 2018.

\bibitem{three_seven}
Z.~Cao and C.-T. Lin, ``Inherent fuzzy entropy for the improvement of eeg
  complexity evaluation,'' \emph{IEEE Transactions on Fuzzy Systems}, vol.~26,
  no.~2, pp. 1032--1035, 2017.

\bibitem{three_eight}
H.-B. Xie, Y.-P. Zheng, J.-Y. Guo, and X.~Chen, ``Cross-fuzzy entropy: A new
  method to test pattern synchrony of bivariate time series,''
  \emph{Information Sciences}, vol. 180, no.~9, pp. 1715--1724, 2010.

\bibitem{four_zero}
Y.~Li, M.~Xu, H.~Zhao, and W.~Huang, ``Hierarchical fuzzy entropy and improved
  support vector machine based binary tree approach for rolling bearing fault
  diagnosis,'' \emph{Mechanism and Machine Theory}, vol.~98, pp. 114--132,
  2016.

\bibitem{four_one}
C.~Liu, K.~Li, L.~Zhao, F.~Liu, D.~Zheng, C.~Liu, and S.~Liu, ``Analysis of
  heart rate variability using fuzzy measure entropy,'' \emph{Computers in
  biology and Medicine}, vol.~43, no.~2, pp. 100--108, 2013.

\bibitem{asvm}
C.~M. Bishop, \emph{Pattern recognition and machine learning}.\hskip 1em plus
  0.5em minus 0.4em\relax springer, 2006.

\bibitem{four_four}
J.-S. Jang, ``Anfis: adaptive-network-based fuzzy inference system,''
  \emph{IEEE transactions on systems, man, and cybernetics}, vol.~23, no.~3,
  pp. 665--685, 1993.

\bibitem{goanew}
S.~Saremi, S.~Mirjalili, and A.~Lewis, ``Grasshopper optimisation algorithm:
  theory and application,'' \emph{Advances in Engineering Software}, vol. 105,
  pp. 30--47, 2017.

\bibitem{five_two}
J.~Kennedy and R.~Eberhart, ``Particle swarm optimization. inproceedings of
  icnn’95-international conference on neural networks 1995 nov 27 (vol. 4,
  pp. 1942--1948),'' \emph{IEEE. View Article}.

\bibitem{five_three}
M.~Settles and T.~Soule, ``Breeding swarms: a ga/pso hybrid,'' in
  \emph{Proceedings of the 7th annual conference on Genetic and evolutionary
  computation}, 2005, pp. 161--168.

\bibitem{anfga}
L.~F. Pereira, S.~A. Patil, C.~D. Mahadeshwar, I.~Mishra, and L.~D'Souza,
  ``Artifact removal from eeg using anfis-ga,'' in \emph{2016 Online
  International Conference on Green Engineering and Technologies
  (IC-GET)}.\hskip 1em plus 0.5em minus 0.4em\relax IEEE, 2016, pp. 1--6.

\bibitem{anfpso}
U.~Kose and A.~Arslan, ``Forecasting chaotic time series via anfis supported by
  vortex optimization algorithm: Applications on electroencephalogram time
  series,'' \emph{Arabian Journal for Science and Engineering}, vol.~42, no.~8,
  pp. 3103--3114, 2017.

\bibitem{aw3}
A.~Bhattacharyya and R.~B. Pachori, ``A multivariate approach for
  patient-specific eeg seizure detection using empirical wavelet transform,''
  \emph{IEEE Transactions on Biomedical Engineering}, vol.~64, no.~9, pp.
  2003--2015, 2017.

\bibitem{aw4}
A.~R. Hassan and M.~I.~H. Bhuiyan, ``Dual tree complex wavelet transform for
  sleep state identification from single channel electroencephalogram,'' in
  \emph{2015 IEEE International Conference on Telecommunications and Photonics
  (ICTP)}.\hskip 1em plus 0.5em minus 0.4em\relax IEEE, 2015, pp. 1--5.

\bibitem{aw5}
A.~Prochazka, J.~Kukal, and O.~Vysata, ``Wavelet transform use for feature
  extraction and eeg signal segments classification,'' in \emph{2008 3rd
  International symposium on communications, control and signal
  processing}.\hskip 1em plus 0.5em minus 0.4em\relax IEEE, 2008, pp. 719--722.

\bibitem{clrs}
T.~H. Cormen, C.~E. Leiserson, R.~L. Rivest, and C.~Stein, \emph{Introduction
  to algorithms}.\hskip 1em plus 0.5em minus 0.4em\relax MIT press, 2009.

\bibitem{two_seven}
A.~Shoeibi, N.~Ghassemi, R.~Alizadehsani, M.~Rouhani, H.~Hosseini-Nejad,
  A.~Khosravi, M.~Panahiazar, and S.~Nahavandi, ``A comprehensive comparison of
  handcrafted features and convolutional autoencoders for epileptic seizures
  detection in eeg signals,'' \emph{Expert Systems with Applications}, vol.
  163, p. 113788.

\bibitem{four_seven}
P.~Cunningham, ``Dimension reduction,'' in \emph{Machine learning techniques
  for multimedia}.\hskip 1em plus 0.5em minus 0.4em\relax Springer, 2008, pp.
  91--112.

\bibitem{four_eight}
S.~Wold, K.~Esbensen, and P.~Geladi, ``Principal component analysis,''
  \emph{Chemometrics and intelligent laboratory systems}, vol.~2, no. 1-3, pp.
  37--52, 1987.

\bibitem{four_nine}
H.~Abdi and L.~J. Williams, ``Principal component analysis,'' \emph{Wiley
  interdisciplinary reviews: computational statistics}, vol.~2, no.~4, pp.
  433--459, 2010.

\bibitem{four_two}
T.~Wen and Z.~Zhang, ``Deep convolution neural network and autoencoders-based
  unsupervised feature learning of eeg signals,'' \emph{IEEE Access}, vol.~6,
  pp. 25\,399--25\,410, 2018.

\bibitem{four_three}
I.~Goodfellow, Y.~Bengio, and A.~Courville, \emph{Deep learning}.\hskip 1em
  plus 0.5em minus 0.4em\relax Cambridge: MIT press, 2016, vol.~1.

\bibitem{rotten2}
H.~Mahami, N.~Ghassemi, M.~T. Darbandy, A.~Shoeibi, S.~Hussain, F.~Nasirzadeh,
  R.~Alizadehsani, D.~Nahavandi, A.~Khosravi, and S.~Nahavandi, ``Material
  recognition for automated progress monitoring using deep learning methods,''
  \emph{arXiv preprint arXiv:2006.16344}, 2020.

\bibitem{five_zero}
Y.~Wang, J.~Liu, J.~Mi{\v{s}}i{\'c}, V.~B. Mi{\v{s}}i{\'c}, S.~Lv, and
  X.~Chang, ``Assessing optimizer impact on dnn model sensitivity to
  adversarial examples,'' \emph{IEEE Access}, vol.~7, pp. 152\,766--152\,776,
  2019.

\bibitem{five_one}
S.~K. Roy, S.~Manna, S.~R. Dubey, and B.~B. Chaudhuri, ``Lisht: Non-parametric
  linearly scaled hyperbolic tangent activation function for neural networks,''
  \emph{arXiv preprint arXiv:1901.05894}, 2019.

\bibitem{astat}
B.~Efron and T.~Hastie, \emph{Computer age statistical inference}.\hskip 1em
  plus 0.5em minus 0.4em\relax Cambridge University Press, 2016, vol.~5.

\bibitem{adt}
J.~Ali, R.~Khan, N.~Ahmad, and I.~Maqsood, ``Random forests and decision
  trees,'' \emph{International Journal of Computer Science Issues (IJCSI)},
  vol.~9, no.~5, p. 272, 2012.

\bibitem{aw1}
M.~Mohammadpoor, A.~Shoeibi, H.~Shojaee \emph{et~al.}, ``A hierarchical
  classification method for breast tumor detection,'' \emph{Iranian Journal of
  Medical Physics}, vol.~13, no.~4, pp. 261--268, 2016.

\bibitem{afuz}
L.~Kuncheva, \emph{Fuzzy classifier design}.\hskip 1em plus 0.5em minus
  0.4em\relax Springer Science \& Business Media, 2000, vol.~49.

\bibitem{amlp}
Y.~LeCun, Y.~Bengio, and G.~Hinton, ``Deep learning,'' \emph{nature}, vol. 521,
  no. 7553, pp. 436--444, 2015.

\bibitem{aw2}
N.~Ghassemi, A.~Shoeibi, and M.~Rouhani, ``Deep neural network with generative
  adversarial networks pre-training for brain tumor classification based on mr
  images,'' \emph{Biomedical Signal Processing and Control}, vol.~57, p.
  101678, 2020.

\bibitem{keras}
F.~Chollet \emph{et~al.}, ``Keras,'' 2015.

\bibitem{nb7}
Y.~Zhang, R.~Yang, and W.~Zhou, ``Roughness-length-based characteristic
  analysis of intracranial eeg and epileptic seizure prediction.''
  \emph{International Journal of Neural Systems}, pp. 2\,050\,072--2\,050\,072,
  2020.

\bibitem{nb8}
Z.~Yu, W.~Zhou, F.~Zhang, F.~Xu, S.~Yuan, Y.~Leng, Y.~Li, and Q.~Yuan,
  ``Automatic seizure detection based on kernel robust probabilistic
  collaborative representation,'' \emph{Medical \& biological engineering \&
  computing}, vol.~57, no.~1, pp. 205--219, 2019.

\bibitem{nb9}
Q.~Yuan, W.~Zhou, L.~Zhang, F.~Zhang, F.~Xu, Y.~Leng, D.~Wei, and M.~Chen,
  ``Epileptic seizure detection based on imbalanced classification and wavelet
  packet transform,'' \emph{Seizure}, vol.~50, pp. 99--108, 2017.

\bibitem{five_four}
D.~Kukolj, ``Design of adaptive takagi--sugeno--kang fuzzy models,''
  \emph{Applied Soft Computing}, vol.~2, no.~2, pp. 89--103, 2002.

\bibitem{five_five}
J.~C. Bezdek, R.~Ehrlich, and W.~Full, ``Fcm: The fuzzy c-means clustering
  algorithm,'' \emph{Computers \& Geosciences}, vol.~10, no. 2-3, pp. 191--203,
  1984.

\bibitem{five_six}
A.~Shoeibi, N.~Ghassemi, H.~Hosseini-Nejad, and M.~Rouhani, ``An efficient
  brain mr images segmentation hardware using kernel fuzzy c-means,'' in
  \emph{2019 26th National and 4th International Iranian Conference on
  Biomedical Engineering (ICBME)}.\hskip 1em plus 0.5em minus 0.4em\relax IEEE,
  2019, pp. 93--99.

\bibitem{newer1}
H.~Leung and S.~Haykin, ``The complex backpropagation algorithm,'' \emph{IEEE
  Transactions on signal processing}, vol.~39, no.~9, pp. 2101--2104, 1991.

\bibitem{newer2}
R.~Rojas, ``The backpropagation algorithm,'' in \emph{Neural networks}.\hskip
  1em plus 0.5em minus 0.4em\relax Springer, 1996, pp. 149--182.

\bibitem{newer3}
M.~Rezakazemi, A.~Dashti, M.~Asghari, and S.~Shirazian, ``H2-selective mixed
  matrix membranes modeling using anfis, pso-anfis, ga-anfis,''
  \emph{International Journal of Hydrogen Energy}, vol.~42, no.~22, pp.
  15\,211--15\,225, 2017.

\bibitem{newer4}
Z.~Ceylan, E.~Pekel, S.~Ceylan, S.~Bulkan \emph{et~al.}, ``Biomass higher
  heating value prediction analysis by anfis, pso-anfis and ga-anfis,''
  \emph{Global Nest Journal}, vol.~20, no.~3, pp. 589--597, 2018.

\bibitem{newer5}
M.~Mosavi, A.~Ayatollahi, and S.~Afrakhteh, ``An efficient method for
  classifying motor imagery using cpso-trained anfis prediction,''
  \emph{Evolving Systems}, pp. 1--18, 2019.

\bibitem{nine}
A.~Nishad and R.~B. Pachori, ``Classification of epileptic electroencephalogram
  signals using tunable-q wavelet transform based filter-bank,'' \emph{Journal
  of Ambient Intelligence and Humanized Computing}, 01 2020.

\bibitem{ten}
E.~Aydemir, T.~Tuncer, and S.~Dogan, ``A tunable-q wavelet transform and
  quadruple symmetric pattern based eeg signal classification method,''
  \emph{Medical Hypotheses}, vol. 134, p. 109519, 01 2020.

\bibitem{one_two}
E.~A. Abdel-Ghaffar, ``Effect of tuning tqwt parameters on epileptic seizure
  detection from eeg signals,'' \emph{2017 12th International Conference on
  Computer Engineering and Systems (ICCES)}, 12 2017.

\bibitem{one_thre}
A.~Bhattacharyya, R.~Pachori, A.~Upadhyay, and U.~Acharya, ``Tunable-q wavelet
  transform based multiscale entropy measure for automated classification of
  epileptic eeg signals,'' \emph{Applied Sciences}, vol.~7, p. 385, 04 2017.

\bibitem{one_five}
A.~R. Hassan, S.~Siuly, and Y.~Zhang, ``Epileptic seizure detection in eeg
  signals using tunable-q factor wavelet transform and bootstrap aggregating,''
  \emph{Computer methods and programs in biomedicine}, vol. 137, pp. 247--259,
  2016.

\bibitem{one_six}
G.~Ravi Shankar~Reddy and R.~Rao, ``Automated identification system for seizure
  eeg signals using tunable-q wavelet transform,'' \emph{Engineering Science
  and Technology, an International Journal}, vol.~20, pp. 1486--1493, 10 2017.

\bibitem{one_eight}
S.~Chen, X.~Zhang, L.~Chen, and Z.~Yang, ``Automatic diagnosis of epileptic
  seizure in electroencephalography signals using nonlinear dynamics
  features,'' \emph{IEEE Access}, vol.~7, pp. 61\,046--61\,056, 2019.

\bibitem{one_nine}
T.~Tuncer, S.~Dogan, and E.~Akbal, ``A novel local senary pattern based
  epilepsy diagnosis system using eeg signals,'' \emph{Australasian Physical \&
  Engineering Sciences in Medicine}, vol.~42, pp. 939--948, 09 2019.

\bibitem{two_zero}
V.~Gupta and R.~B. Pachori, ``Epileptic seizure identification using entropy of
  fbse based eeg rhythms,'' \emph{Biomedical Signal Processing and Control},
  vol.~53, p. 101569, 08 2019.

\bibitem{two_one}
S.~Raghu, N.~Sriraam, A.~S. Hegde, and P.~L. Kubben, ``A novel approach for
  classification of epileptic seizures using matrix determinant,'' \emph{Expert
  Systems with Applications}, vol. 127, pp. 323--341, 08 2019.

\bibitem{two_two}
R.~R. Sharma, P.~Varshney, R.~B. Pachori, and S.~K. Vishvakarma, ``Automated
  system for epileptic eeg detection using iterative filtering,'' \emph{IEEE
  Sensors Letters}, vol.~2, pp. 1--4, 12 2018.

\bibitem{two_three}
A.~K. Tiwari, R.~B. Pachori, V.~Kanhangad, and B.~K. Panigrahi, ``Automated
  diagnosis of epilepsy using key-point-based local binary pattern of eeg
  signals,'' \emph{IEEE Journal of Biomedical and Health Informatics}, vol.~21,
  pp. 888--896, 07 2017.

\bibitem{two_four}
K.~D. Tzimourta, A.~T. Tzallas, N.~Giannakeas, L.~G. Astrakas, D.~G.
  Tsalikakis, P.~Angelidis, and M.~G. Tsipouras, ``A robust methodology for
  classification of epileptic seizures in eeg signals,'' \emph{Health and
  Technology}, vol.~9, pp. 135--142, 09 2018.

\bibitem{a1}
D.~Lu and J.~Triesch, ``Residual deep convolutional neural network for eeg
  signal classification in epilepsy,'' \emph{arXiv preprint arXiv:1903.08100},
  2019.

\bibitem{a3}
M.~U. Abbasi, A.~Rashad, A.~Basalamah, and M.~Tariq, ``Detection of epilepsy
  seizures in neo-natal eeg using lstm architecture,'' \emph{IEEE Access},
  vol.~7, pp. 179\,074--179\,085, 2019.

\bibitem{a4}
A.~M. Karim, {\"O}.~Karal, and F.~{\c{C}}elebi, ``A new automatic epilepsy
  serious detection method by using deep learning based on discrete wavelet
  transform,'' \emph{no}, vol.~4, pp. 15--18, 2018.

\bibitem{a5}
T.~Wen and Z.~Zhang, ``Deep convolution neural network and autoencoders-based
  unsupervised feature learning of eeg signals,'' \emph{IEEE Access}, vol.~6,
  pp. 25\,399--25\,410, 2018.

\bibitem{a6}
R.~Abiyev, M.~Arslan, J.~B. Idoko, B.~Sekeroglu, and A.~Ilhan, ``Identification
  of epileptic eeg signals using convolutional neural networks,'' \emph{Applied
  Sciences}, vol.~10, no.~12, p. 4089, 2020.

\bibitem{a7}
N.~Ilakiyaselvan, A.~N. Khan, and A.~Shahina, ``Deep learning approach to
  detect seizure using reconstructed phase space images,'' \emph{Journal of
  Biomedical Research}, vol.~34, no.~3, p. 240, 2020.

\bibitem{a9}
A.~Singh, N.~Pusarla, S.~Sharma, and T.~Kumar, ``Cnn-based epilepsy detection
  using image like features of eeg signals,'' in \emph{2020 International
  Conference on Electrical and Electronics Engineering (ICE3)}.\hskip 1em plus
  0.5em minus 0.4em\relax IEEE, 2020, pp. 280--284.

\bibitem{a10}
X.~Gao, X.~Yan, P.~Gao, X.~Gao, and S.~Zhang, ``Automatic detection of
  epileptic seizure based on approximate entropy, recurrence quantification
  analysis and convolutional neural networks,'' \emph{Artificial Intelligence
  in Medicine}, vol. 102, p. 101711, 2020.

\bibitem{a11}
J.~Lian, Y.~Zhang, R.~Luo, G.~Han, W.~Jia, and C.~Li, ``Pair-wise matching of
  eeg signals for epileptic identification via convolutional neural network,''
  \emph{IEEE Access}, vol.~8, pp. 40\,008--40\,017, 2020.

\bibitem{a12}
K.~Akyol, ``Stacking ensemble based deep neural networks modeling for effective
  epileptic seizure detection,'' \emph{Expert Systems with Applications}, vol.
  148, p. 113239, 2020.

\bibitem{di2}
I.~B. Slimen and H.~Seddik, ``Automatic recognition of epileptiform eeg
  abnormalities using machine learning approaches,'' in \emph{2020 5th
  International Conference on Advanced Technologies for Signal and Image
  Processing (ATSIP)}.\hskip 1em plus 0.5em minus 0.4em\relax IEEE, 2020, pp.
  1--4.

\bibitem{di3}
A.~Sharmila and P.~Geethanjali, ``Evaluation of time domain features on
  detection of epileptic seizure from eeg signals,'' \emph{Health and
  Technology}, vol.~10, no.~3, pp. 711--722, 2020.

\bibitem{di4}
H.~D. Praveena, C.~Subhas, and K.~R. Naidu, ``Detection of epileptic seizure
  based on relieff algorithm and multi-support vector machine,'' in
  \emph{Cognitive Informatics and Soft Computing}.\hskip 1em plus 0.5em minus
  0.4em\relax Springer, 2020, pp. 13--28.

\bibitem{na1}
M.~Z. Abedin, S.~Akther, and M.~S. Hossain, ``An artificial neural network
  model for epilepsy seizure detection,'' in \emph{2019 5th International
  Conference on Advances in Electrical Engineering (ICAEE)}.\hskip 1em plus
  0.5em minus 0.4em\relax IEEE, 2019, pp. 860--865.

\bibitem{na5}
N.~Parveen and S.~H. Saeed, ``Tunable-q wavelet transform based entropy
  measures for identification of epileptic seizures,'' \emph{European Journal
  of Molecular \& Clinical Medicine}, vol.~7, no.~11, pp. 4702--4717, 2021.

\bibitem{na7}
G.~C. Jana, A.~Sabath, and A.~Agrawal, ``Performance analysis of supervised
  machine learning algorithms for epileptic seizure detection with high
  variability eeg datasets: A comparative study,'' in \emph{2019 International
  Conference on Electrical, Electronics and Computer Engineering
  (UPCON)}.\hskip 1em plus 0.5em minus 0.4em\relax IEEE, 2019, pp. 1--6.

\bibitem{na8}
S.~Ramakrishnan and A.~M. Murugavel, ``Epileptic seizure detection using
  fuzzy-rules-based sub-band specific features and layered multi-class svm,''
  \emph{Pattern Analysis and Applications}, vol.~22, no.~3, pp. 1161--1176,
  2019.

\bibitem{na10}
M.~Thilagaraj, M.~P. Rajasekaran, and N.~A. Kumar, ``Tsallis entropy: as a new
  single feature with the least computation time for classification of
  epileptic seizures,'' \emph{Cluster Computing}, vol.~22, no.~6, pp.
  15\,213--15\,221, 2019.

\bibitem{na12}
H.~Al-Hadeethi, S.~Abdulla, M.~Diykh, R.~C. Deo, and J.~H. Green, ``Adaptive
  boost ls-svm classification approach for time-series signal classification in
  epileptic seizure diagnosis applications,'' \emph{Expert Systems with
  Applications}, vol. 161, p. 113676, 2020.

\bibitem{na13}
G.~Singh, M.~Kaur, and B.~Singh, ``Detection of epileptic seizure eeg signal
  using multiscale entropies and complete ensemble empirical mode
  decomposition,'' \emph{Wireless Personal Communications}, vol. 116, no.~1,
  pp. 845--864, 2021.

\bibitem{na14}
M.~B. Qureshi, M.~Afzaal, M.~S. Qureshi, M.~Fayaz \emph{et~al.}, ``Machine
  learning-based eeg signals classification model for epileptic seizure
  detection,'' \emph{Multimedia Tools and Applications}, pp. 1--29, 2021.

\bibitem{na17}
M.~R. Kumar and Y.~S. Rao, ``Epileptic seizures classification in eeg signal
  based on semantic features and variational mode decomposition,''
  \emph{Cluster Computing}, vol.~22, no.~6, pp. 13\,521--13\,531, 2019.

\bibitem{na18}
J.~Wu, T.~Zhou, and T.~Li, ``Detecting epileptic seizures in eeg signals with
  complementary ensemble empirical mode decomposition and extreme gradient
  boosting,'' \emph{Entropy}, vol.~22, no.~2, p. 140, 2020.

\bibitem{na19}
S.~Ashokkumar, G.~MohanBabu, and S.~Anupallavi, ``A novel two-band equilateral
  wavelet filter bank method for an automated detection of seizure from eeg
  signals,'' \emph{International Journal of Imaging Systems and Technology},
  vol.~30, no.~4, pp. 978--993, 2020.

\bibitem{na20}
S.~Raghu, N.~Sriraam, Y.~Temel, S.~V. Rao, A.~S. Hegde, and P.~L. Kubben,
  ``Performance evaluation of dwt based sigmoid entropy in time and frequency
  domains for automated detection of epileptic seizures using svm classifier,''
  \emph{Computers in biology and medicine}, vol. 110, pp. 127--143, 2019.

\bibitem{na21}
D.~K. Atal and M.~Singh, ``A hybrid feature extraction and machine learning
  approaches for epileptic seizure detection,'' \emph{Multidimensional Systems
  and Signal Processing}, pp. 1--23, 2019.

\bibitem{na27}
Y.~Liu, B.~Jiang, J.~Feng, J.~Hu, and H.~Zhang, ``Classification of eeg signals
  for epileptic seizures using feature dimension reduction algorithm based on
  lpp,'' \emph{Multimedia Tools and Applications}, pp. 1--22, 2020.

\bibitem{na28}
M.~G. Tsipouras, ``Spectral information of eeg signals with respect to epilepsy
  classification,'' \emph{EURASIP Journal on Advances in Signal Processing},
  vol. 2019, no.~1, pp. 1--17, 2019.

\bibitem{na29}
H.-S. Chiang, M.-Y. Chen, and Y.-J. Huang, ``Wavelet-based eeg processing for
  epilepsy detection using fuzzy entropy and associative petri net,''
  \emph{IEEE Access}, vol.~7, pp. 103\,255--103\,262, 2019.

\bibitem{na30}
I.~Aliyu and C.~G. Lim, ``Selection of optimal wavelet features for epileptic
  eeg signal classification with lstm,'' \emph{Neural Computing and
  Applications}, pp. 1--21, 2021.

\bibitem{na31}
M.~Sameer and B.~Gupta, ``Beta band as a biomarker for classification between
  interictal and ictal states of epileptical patients,'' in \emph{2020 7th
  International Conference on Signal Processing and Integrated Networks
  (SPIN)}.\hskip 1em plus 0.5em minus 0.4em\relax IEEE, 2020, pp. 567--570.

\bibitem{nb2}
K.~Tzimourta, A.~Tzallas, N.~Giannakeas, L.~Astrakas, D.~Tsalikakis, and
  M.~Tsipouras, ``Epileptic seizures classification based on long-term eeg
  signal wavelet analysis,'' in \emph{International Conference on Biomedical
  and Health Informatics}.\hskip 1em plus 0.5em minus 0.4em\relax Springer,
  2017, pp. 165--169.

\bibitem{nb3}
K.~D. Tzimourta, A.~T. Tzallas, N.~Giannakeas, L.~G. Astrakas, D.~G.
  Tsalikakis, P.~Angelidis, and M.~G. Tsipouras, ``A robust methodology for
  classification of epileptic seizures in eeg signals,'' \emph{Health and
  Technology}, vol.~9, no.~2, pp. 135--142, 2019.

\bibitem{nb4}
N.~D. Truong and O.~Kavehei, ``Low precision electroencephalogram for seizure
  detection with convolutional neural network,'' in \emph{2019 IEEE
  International Conference on Artificial Intelligence Circuits and Systems
  (AICAS)}.\hskip 1em plus 0.5em minus 0.4em\relax IEEE, 2019, pp. 299--301.

\bibitem{nb5}
M.~Geng, W.~Zhou, G.~Liu, C.~Li, and Y.~Zhang, ``Epileptic seizure detection
  based on stockwell transform and bidirectional long short-term memory,''
  \emph{Ieee Transactions on Neural Systems and Rehabilitation Engineering},
  vol.~28, no.~3, pp. 573--580, 2020.

\bibitem{nb6}
M.~Nassralla, M.~Haidar, H.~Alawieh, A.~El~Hajj, and Z.~Dawy, ``Patient-aware
  eeg-based feature and classifier selection for e-health epileptic seizure
  prediction,'' in \emph{2018 IEEE Global Communications Conference
  (GLOBECOM)}.\hskip 1em plus 0.5em minus 0.4em\relax IEEE, 2018, pp. 1--6.

\bibitem{nb10}
M.~Zhou, C.~Tian, R.~Cao, B.~Wang, Y.~Niu, T.~Hu, H.~Guo, and J.~Xiang,
  ``Epileptic seizure detection based on eeg signals and cnn,'' \emph{Frontiers
  in neuroinformatics}, vol.~12, p.~95, 2018.

\bibitem{nb11}
N.~Mahmoodian, A.~Boese, M.~Friebe, and J.~Haddadnia, ``Epileptic seizure
  detection using cross-bispectrum of electroencephalogram signal,''
  \emph{seizure}, vol.~66, pp. 4--11, 2019.

\bibitem{nb12}
J.~Martinez-del Rincon, M.~J. Santofimia, X.~del Toro, J.~Barba, F.~Romero,
  P.~Navas, and J.~C. Lopez, ``Non-linear classifiers applied to eeg analysis
  for epilepsy seizure detection,'' \emph{Expert Systems with Applications},
  vol.~86, pp. 99--112, 2017.

\bibitem{nb13}
Q.~Zhan and W.~Hu, ``An epilepsy detection method using multiview clustering
  algorithm and deep features,'' \emph{Computational and Mathematical Methods
  in Medicine}, vol. 2020, 2020.

\bibitem{nb14}
G.~Liu, W.~Zhou, and M.~Geng, ``Automatic seizure detection based on
  s-transform and deep convolutional neural network,'' \emph{International
  journal of neural systems}, vol.~30, no.~04, p. 1950024, 2020.

\bibitem{nb15}
S.~T. Jaafar and M.~Mohammadi, ``Epileptic seizure detection using deep
  learning approach,'' \emph{UHD Journal of Science and Technology}, vol.~3,
  no.~2, pp. 41--50, 2019.

\bibitem{nb16}
Y.~Li, Z.~Yu, Y.~Chen, C.~Yang, Y.~Li, X.~Allen~Li, and B.~Li, ``Automatic
  seizure detection using fully convolutional nested lstm,''
  \emph{International journal of neural systems}, vol.~30, no.~04, p. 2050019,
  2020.

\bibitem{nb17}
N.~D. Truong, A.~D. Nguyen, L.~Kuhlmann, M.~R. Bonyadi, J.~Yang, S.~Ippolito,
  and O.~Kavehei, ``Integer convolutional neural network for seizure
  detection,'' \emph{IEEE Journal on Emerging and Selected Topics in Circuits
  and Systems}, vol.~8, no.~4, pp. 849--857, 2018.

\bibitem{nb18}
E.~Alickovic, J.~Kevric, and A.~Subasi, ``Performance evaluation of empirical
  mode decomposition, discrete wavelet transform, and wavelet packed
  decomposition for automated epileptic seizure detection and prediction,''
  \emph{Biomedical signal processing and control}, vol.~39, pp. 94--102, 2018.

\bibitem{nb19}
Q.~Yuan, W.~Zhou, Y.~Liu, and J.~Wang, ``Epileptic seizure detection with
  linear and nonlinear features,'' \emph{Epilepsy \& Behavior}, vol.~24, no.~4,
  pp. 415--421, 2012.

\bibitem{nb20}
B.~Abbaszadeh and M.~C. Yagoub, ``Optimum window size and overlap for robust
  probabilistic prediction of seizures with ieeg,'' in \emph{2019 IEEE
  Conference on Computational Intelligence in Bioinformatics and Computational
  Biology (CIBCB)}.\hskip 1em plus 0.5em minus 0.4em\relax IEEE, 2019, pp.
  1--5.

\bibitem{nb21}
M.~Mohammadi, N.~A. Khan, and A.~A. Pouyan, ``Automatic seizure detection using
  a highly adaptive directional time--frequency distribution,''
  \emph{Multidimensional Systems and Signal Processing}, vol.~29, no.~4, pp.
  1661--1678, 2018.

\bibitem{nb22}
Z.~Yu, Y.~Li, Q.~Yuan, and W.~Zhou, ``Epileptic seizure detection based on
  local mean decomposition and dictionary pair learning,'' in \emph{2018
  International Conference on Information Systems and Computer Aided Education
  (ICISCAE)}.\hskip 1em plus 0.5em minus 0.4em\relax IEEE, 2018, pp. 432--435.

\bibitem{nb23}
C.~Sun, H.~Cui, W.~Zhou, W.~Nie, X.~Wang, and Q.~Yuan, ``Epileptic seizure
  detection with eeg textural features and imbalanced classification based on
  easyensemble learning,'' \emph{International journal of neural systems},
  vol.~29, no.~10, p. 1950021, 2019.

\bibitem{nb24}
A.~Abugabah, A.~A. AlZubi, M.~Al-Maitah, and A.~Alarifi, ``Brain epilepsy
  seizure detection using bio-inspired krill herd and artificial alga optimized
  neural network approaches,'' \emph{Journal of Ambient Intelligence and
  Humanized Computing}, vol.~12, no.~3, pp. 3317--3328, 2021.

\bibitem{nb25}
J.~Kevric and A.~Subasi, ``The effect of multiscale pca de-noising in epileptic
  seizure detection,'' \emph{Journal of medical systems}, vol.~38, no.~10, pp.
  1--13, 2014.

\bibitem{nb26}
X.~Ma, N.~Yu, and W.~Zhou, ``Using dictionary pair learning for seizure
  detection,'' \emph{International journal of neural systems}, vol.~29, no.~04,
  p. 1850005, 2019.

\bibitem{nb27}
H.~Niknazar, S.~R. Mousavi, M.~Niknazar, V.~Mardanlou, and B.~N. Coelho,
  ``Performance analysis of eeg seizure detection features,'' \emph{Epilepsy
  Research}, vol. 167, p. 106483, 2020.

\bibitem{rrt5}
S.~K. Dhull, K.~K. Singh \emph{et~al.}, ``A review on automatic epilepsy
  detection from eeg signals,'' in \emph{Advances in Communication and
  Computational Technology}.\hskip 1em plus 0.5em minus 0.4em\relax Springer,
  2021, pp. 1441--1454.

\bibitem{rrt1}
M.~K. Siddiqui, R.~Morales-Menendez, X.~Huang, and N.~Hussain, ``A review of
  epileptic seizure detection using machine learning classifiers,'' \emph{Brain
  informatics}, vol.~7, pp. 1--18, 2020.

\bibitem{rrt2}
P.~Boonyakitanont, A.~Lek-Uthai, K.~Chomtho, and J.~Songsiri, ``A review of
  feature extraction and performance evaluation in epileptic seizure detection
  using eeg,'' \emph{Biomedical Signal Processing and Control}, vol.~57, p.
  101702, 2020.

\bibitem{rrt3}
L.~Boubchir, B.~Daachi, and V.~Pangracious, ``A review of feature extraction
  for eeg epileptic seizure detection and classification,'' in \emph{2017 40th
  International Conference on Telecommunications and Signal Processing
  (TSP)}.\hskip 1em plus 0.5em minus 0.4em\relax IEEE, 2017, pp. 456--460.

\bibitem{rrt4}
F.~S. Leijten, D.~T. Consortium, J.~van Andel, C.~Ungureanu, J.~Arends, F.~Tan,
  J.~van Dijk, G.~Petkov, S.~Kalitzin, T.~Gutter \emph{et~al.}, ``Multimodal
  seizure detection: a review,'' \emph{Epilepsia}, vol.~59, pp. 42--47, 2018.

\bibitem{rtltrt1}
R.~Alizadehsani, D.~Sharifrazi, N.~H. Izadi, J.~H. Joloudari, A.~Shoeibi, J.~M.
  Gorriz, S.~Hussain, J.~E. Arco, Z.~A. Sani, F.~Khozeimeh \emph{et~al.},
  ``Uncertainty-aware semi-supervised method using large unlabelled and limited
  labeled covid-19 data,'' \emph{arXiv preprint arXiv:2102.06388}, 2021.

\bibitem{five_seven}
A.~T. Tzallas, M.~G. Tsipouras, D.~G. Tsalikakis, E.~C. Karvounis, L.~Astrakas,
  S.~Konitsiotis, and M.~Tzaphlidou, ``Automated epileptic seizure detection
  methods: a review study,'' \emph{Epilepsy-histological,
  electroencephalographic and psychological aspects}, pp. 75--98, 2012.

\bibitem{five_eight}
A.~F. Hussein, N.~Arunkumar, C.~Gomes, A.~K. Alzubaidi, Q.~A. Habash,
  L.~Santamaria-Granados, J.~F. Mendoza-Moreno, and G.~Ramirez-Gonzalez,
  ``Focal and non-focal epilepsy localization: A review,'' \emph{IEEE Access},
  vol.~6, pp. 49\,306--49\,324, 2018.

\bibitem{four_six}
U.~R. Acharya, S.~V. Sree, G.~Swapna, R.~J. Martis, and J.~S. Suri, ``Automated
  eeg analysis of epilepsy: a review,'' \emph{Knowledge-Based Systems},
  vol.~45, pp. 147--165, 2013.

\bibitem{rro1}
T.~N. Sainath, A.~Narayanan, R.~J. Weiss, E.~Variani, K.~W. Wilson,
  M.~Bacchiani, and I.~Shafran, ``Reducing the computational complexity of
  multimicrophone acoustic models with integrated feature extraction,'' 2016.

\bibitem{rro2}
A.~Humeau-Heurtier, ``Texture feature extraction methods: A survey,''
  \emph{IEEE Access}, vol.~7, pp. 8975--9000, 2019.

\bibitem{rro3}
M.~Blanco-Velasco, F.~Cruz-Rold{\'a}n, F.~L{\'o}pez-Ferreras, A.~Bravo-Santos,
  and D.~Martinez-Munoz, ``A low computational complexity algorithm for ecg
  signal compression,'' \emph{Medical engineering \& physics}, vol.~26, no.~7,
  pp. 553--568, 2004.

\bibitem{rro4}
C.~Zhou and A.~Wieser, ``Modified jaccard index analysis and adaptive feature
  selection for location fingerprinting with limited computational
  complexity,'' \emph{Journal of Location Based Services}, vol.~13, no.~2, pp.
  128--157, 2019.

\bibitem{rro5}
T.~Chen, E.~Mazomenos, K.~Maharatna, S.~Dasmahapatra, and M.~Niranjan, ``On the
  trade-off of accuracy and computational complexity for classifying normal and
  abnormal ecg in remote cvd monitoring systems,'' in \emph{2012 IEEE Workshop
  on Signal Processing Systems}.\hskip 1em plus 0.5em minus 0.4em\relax IEEE,
  2012, pp. 37--42.

\bibitem{rtltrt2}
R.~Alizadehsani, M.~Roshanzamir, S.~Hussain, A.~Khosravi, A.~Koohestani, M.~H.
  Zangooei, M.~Abdar, A.~Beykikhoshk, A.~Shoeibi, A.~Zare \emph{et~al.},
  ``Handling of uncertainty in medical data using machine learning and
  probability theory techniques: A review of 30 years (1991--2020),''
  \emph{Annals of Operations Research}, pp. 1--42, 2021.

\bibitem{newer6}
C.~Chen, R.~John, J.~Twycross, and J.~M. Garibaldi, ``Type-1 and interval
  type-2 anfis: A comparison,'' in \emph{2017 IEEE International Conference on
  Fuzzy Systems (FUZZ-IEEE)}.\hskip 1em plus 0.5em minus 0.4em\relax IEEE,
  2017, pp. 1--6.

\bibitem{newer7}
E.-H. Kim, S.-K. Oh, and W.~Pedrycz, ``Design of reinforced interval type-2
  fuzzy c-means-based fuzzy classifier,'' \emph{IEEE Transactions on Fuzzy
  Systems}, vol.~26, no.~5, pp. 3054--3068, 2017.

\bibitem{newer8}
O.~Carvajal, P.~Melin, I.~Miramontes, and G.~Prado-Arechiga, ``Optimal design
  of a general type-2 fuzzy classifier for the pulse level and its hardware
  implementation,'' \emph{Engineering Applications of Artificial Intelligence},
  vol.~97, p. 104069.

\bibitem{newer9}
N.~S. Bajestani, A.~V. Kamyad, E.~N. Esfahani, and A.~Zare, ``Prediction of
  retinopathy in diabetic patients using type-2 fuzzy regression model,''
  \emph{European Journal of Operational Research}, vol. 264, no.~3, pp.
  859--869, 2018.

\bibitem{newer10}
N.~S. Bajestani, A.~V. Kamyad, and A.~Zare, ``A piecewise type-2 fuzzy
  regression model,'' \emph{International Journal of Computational Intelligence
  Systems}, vol.~10, no.~1, pp. 734--744, 2017.

\bibitem{newer11}
M.~Soltani, A.~J. Telmoudi, L.~Chaouech, M.~Ali, and A.~Chaari, ``Design of a
  robust interval-valued type-2 fuzzy c-regression model for a nonlinear system
  with noise and outliers,'' \emph{Soft Computing}, vol.~23, no.~15, pp.
  6125--6134, 2019.

\end{thebibliography}

\end{document}